# Improved sensitivity and quantification for $^{29}$Si NMR experiments on solids using UDEFT (Uniform Driven Equilibrium Fourier Transform)

Nghia Tuan Duong,[1,#] Julien Trébosc,[1,2*] Olivier Lafon,[1,3] Jean-Paul Amoureux[1,4*]

[1] Univ. Lille, Centrale Lille, ENSCL, Univ. Artois, CNRS-8181, UCCS – Unit of Catalysis and Chemistry of Solids, F-59000 Lille, France.
[2] Univ. Lille, CNRS-FR2638, Fédération Chevreul, F-59000 Lille, France.
[3] Institut Universitaire de France, 1 rue Descartes, F-75231 Paris, France.
[4] Bruker Biospin, 34 rue de l'industrie, F-67166 Wissembourg, France.
[#] Present address: RIKEN-JEOL Collaboration Center, Yokohama, Kanagawa 230-0045, Japan.

Email addresses: julien.trebosc@univ-lille.fr   jean-paul.amoureux@univ-lille.fr

**Abstract.** We demonstrate the possibility to use UDEFT (Uniform Driven Equilibrium Fourier Transform) technique in order to improve the sensitivity and the quantification of one-dimensional $^{29}$Si NMR experiments under Magic-Angle Spinning (MAS). We derive an analytical expression of the signal-to-noise ratios of UDEFT and single-pulse (SP) experiments subsuming the contributions of transient and steady-state regimes. Using numerical spin dynamics simulations and experiments on $^{29}$Si-enriched amorphous silica and borosilicate glass, we show that $59_{180}298_{0}59_{180}$ refocusing composite π-pulse and the adiabatic inversion using tanh/tan modulation improve the robustness of UDEFT technique to rf-inhomogeneity, offset, and chemical shift anisotropy. These pulses combined with a two-step phase cycling limit the pulse imperfections and the artifacts produced by stimulated echoes. The sensitivity of SP, UDEFT and CPMG (Carr-Purcell Meiboom-Gill) techniques are compared experimentally on functionalized and non-functionalized mesoporous silica. Furthermore, experiments on a flame retardant material prove that UDEFT technique provides a better quantification of $^{29}$Si sites with higher sensitivity than SP method.

## I   Introduction

$^{29}$Si NMR spectroscopy has been used for the characterization of a wide range of solids, including silicon alloys [1–4], silicon-containing organic compounds [5,6] and polymers [7], silicon nitride and carbide ceramics [8,9], crystalline and amorphous silicates, including minerals [10-12], zeolites [13,14], cements [15,16], silica-supported catalysts [17-19], acid heterogeneous catalysts made of amorphous aluminosilicates [20] and glasses [21-23]. $^{29}$Si is a spin-1/2 isotope, but its NMR sensitivity is small due to its low natural abundance (4.7 %), its moderate gyromagnetic ratio ($\gamma_{29Si} \approx 0.2\gamma_{1H}$) and its very long longitudinal relaxation times, $T_1$, which can reach tens of hours [24,25].

Various approaches have been proposed to enhance the NMR sensitivity of $^{29}$Si nuclei. The CPMG (Carr-Purcell Meiboom-Gill) sequence [26], which consists of an excitation pulse followed by a train of spin-echoes (Fig.**1d**), allows the acquisition of multiple echoes in every scans [27,28]. These echoes result from the refocusing of the inhomogeneous broadening of $^{29}$Si signal, produced especially by the distribution of isotropic chemical shifts in disordered solids. However, when the flip-angle of the refocusing pulses is distinct from π, they can convert the transverse magnetization into longitudinal one and create stimulated echoes that may pollute the CPMG signal [29,30]. A variant of CPMG, named PIETA (Phase-Incremented Echo-Train Acquisition), has been introduced to suppress the contribution of the stimulated echoes and has been applied to separate isotropic and anisotropic chemical shifts or to measure the $J_{29Si-29Si}$ couplings [30–32].

In the case of sites exhibiting different decay times, CPMG spectra are not directly quantitative. However, quantitative information can then be retrieved by measuring these decay times for the different sites [33]. Nevertheless, this approach is only applicable in the case of high signal-to-noise (*S/N*) ratio.

For protonated solids, another strategy to enhance the $^{29}$Si sensitivity is to transfer the polarization of protons to $^{29}$Si nuclei using CPMAS (Cross-Polarization under Magic-Angle Spinning) sequence [34,35]. The sensitivity gains of this experiment stem from the larger polarization and faster longitudinal relaxation of protons with respect to $^{29}$Si nuclei. Furthermore, the sensitivity of $^{1}$H→$^{29}$Si CPMAS has been further enhanced by using CPMG detection [36]. Recently $^{1}$H→$^{29}$Si multiple-contact CPMAS scheme has also been applied to record quantitative $^{29}$Si NMR spectra [37].

Correspondingly, the sensitivity of $^{29}$Si NMR can also be enhanced by DNP (Dynamic Nuclear Polarization) under MAS, i.e. the microwave-driven transfer of polarization from unpaired electrons to $^{29}$Si nuclei. This approach has been demonstrated first at low static magnetic field $B_0$ = 1.4 T [38], and more recently at $B_0 \geq$ 9.4 T [39–41]. DNP can be combined with either $^{1}$H→$^{29}$Si CPMAS [39], its multiple contact version [41], or CPMG [42].

Here, we propose the use of UDEFT (Uniform Driven Equilibrium Fourier Transform) technique (Fig.**1b**) to enhance the NMR sensitivity of $^{29}$Si on solids, without resorting to polarization transfer. Like CPMG, UDEFT can be applied for non-protonated samples and does not require the additional presence of polarizing agents into the sample. The UDEFT scheme derives from the DEFT (Driven Equilibrium Fourier Transform) sequence (Fig.**1a**) [43]. The DEFT sequence consists of a spin-echo followed by a flip-back pulse, which returns the transverse magnetization to the *z*-axis after the acquisition. This sequence has been applied for the NMR acquisition of slowly relaxing nuclei, such as $^{13}$C or $^{29}$Si, in solutions [44,45]. It has also been used to suppress the signal of water in NMR spectroscopy [46] or to accelerate the acquisition of images in magnetic resonance microscopy and imaging by driving back the magnetization of water toward the *z*-axis [47,48]. However, DEFT is sensitive to radio-frequency (rf) field inhomogeneities and resonance offsets [49–52]. It has been shown that the UDEFT variant using adiabatic π-pulses circumvents this issue and is widely used for the direct excitation of $^{13}$C spectra [53]. Nevertheless, to the best of our knowledge, neither DEFT nor UDEFT techniques have been used so far to acquire NMR spectra of solids.

We demonstrate here that UDEFT enhances the sensitivity for the NMR detection of $^{29}$Si nuclei in solids and improve the quantification of $^{29}$Si sites. We derive the analytical expression of the *S/N* ratio of UDEFT by taking into account the contributions of transient and steady-state signals. We also determine the minimum recycle delay for the acquisition of quantitative UDEFT spectra and their *S/N* ratios. Using numerical simulations and experiments on $^{29}$Si labeled amorphous silica and borosilicate glass, we analyze the robustness to the synchronization with the sample rotation, rf-inhomogeneity, offset and CSA (Chemical Shift Anisotropy). This analysis allows selecting the optimal phases for the pulses and the most robust composite or adiabatic π-pulses to refocus or invert the $^{29}$Si magnetization in the UDEFT sequence. $^{29}$Si experiments on unlabeled mesoporous silica samples are used to compare the sensitivity of UDEFT, single-pulse (SP) and CPMG techniques. SP and UDEFT experiments are also compared to quantify the relative amount of di- (D: $SiO_2C_2$) and tri-functional (T: $SiO_3C$) sites in silicon resin with flame retardant properties.

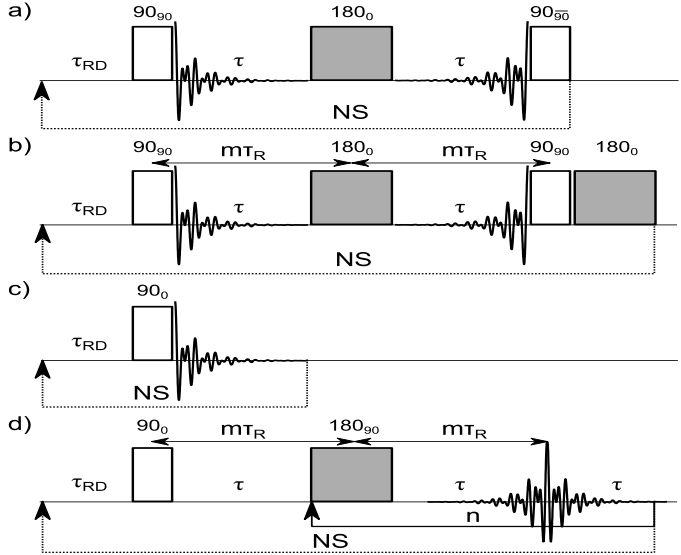

**Fig.1.** Pulse sequences of (a) DEFT, (b) UDEFT, (c) SP, and (d) spin-echo/CPMG experiments. The white and dark rectangles represent the π/2- and π-pulses, respectively. $T_R$ denotes the rotor period, $τ_{RD}$ the recycle delay, τ the echo delay, which is rotor synchronized with respect to the middle to the surrounding pulses. *NS* is the number of scans, and *n* the number of acquired echoes in CPMG. For UDEFT experiments, a two-step phase cycle is employed: the phases of the refocusing and inversion pulses are incremented by 180°, whereas that of the receiver remains constant. The length of each rf element is called $τ_p$ in the following.

## II UDEFT sequence

The original DEFT sequence, shown in Fig.**1a**, can be written as: $90_{90} – τ$ (sampling) $– 180_0 – τ – 90_{-90}$ [43], where τ is the echo delay and $θ_φ$ denotes a rectangular, resonant pulse with flip-angle θ and phase φ, both values given in degrees. The first π/2-pulse flips the magnetization from the z- to the x-axis. The NMR signal is detected during the first τ delay. The central π-pulse refocuses the evolution under the isotropic chemical shifts and the transverse magnetization points toward the x-axis at the end of the second τ delay. Then, the second π/2-pulse returns the magnetization back to the initial z-axis. In the case of ideal pulses and without relaxation, the longitudinal magnetization after the second π/2-pulse is equal to the magnetization at thermal equilibrium and the DEFT scheme can be repeated indefinitely without signal decay. However, experimental pulse imperfections and relaxation lead to magnetization losses and the signal detected during each scan decreases for an increasing number of scans.

The UDEFT sequence shown in Fig.**1b** can be written as $[90_{90} – τ$ (sampling) $– 180_0 – τ – 90_{90}]180_0$ [53]. In UDEFT, both π/2-pulses have identical phases. Such modification improves the robustness to rf-inhomogeneity since for resonant irradiation, the UDEFT sequence mimics the behavior of the $90_{90}180_090_{90}$ composite pulse [54,55] to invert the magnetization. It must be emphasized here that even if the first three pulses of the sequence are those of this $90_{90}180_090_{90}$ composite pulse, the UDEFT sequence by itself does not behave as a composite pulse; it is only inspired by it. However, as will be seen in the following the UDEFT sequence is much more robust than the DEFT one. After the second π/2-pulse, the magnetization points toward the -z direction and a second π-pulse is employed to return the magnetization back to the initial z-axis.

For solution-state experiments, the refocusing element of UDEFT was a composite adiabatic π-pulse made of three smoothed Chirp pulses with relative lengths of $τ_P/4$, $τ_P/2$ and $τ_P/4$ [56], whereas the inversion pulse was a single smoothed Chirp adiabatic pulse [57]. In the case of rotating solids, the CSA produces an additional modulation of the resonance frequency, which can interfere with the sweep of the frequency offset during adiabatic pulses. Consequently, these pulses in solids are only efficient for moderate CSA and MAS frequencies or when they are short with high rf-power [58–60]. Therefore, alternatives to smoothed Chirp adiabatic pulses were tested with refocusing and inversion composite π-pulses and these are listed in Tables S1 and S2, respectively. As refocusing element, we also tested several composite tanh/tan adiabatic pulses, including the BIR-4 ($B_1$-Insensitive Rotation) one [61,62] and three successive such pulses with relative lengths of $\tau_P/4$, $\tau_P/2$ and $\tau_P/4$ [56]. For the inversion element, we also employed the tanh/tan adiabatic pulse, which has been developed to achieve fast broadband inversion and has been applied to solid-state experiments [63,64].

During a tanh/tan pulse with a length $\tau_p$, the instantaneous rf-amplitude is equal to

$$\nu_1(t) = \begin{cases} \nu_{1\max} \cdot \tanh\{2\zeta t/\tau_p\} & 0 \leq t < \tau_p/2 \\ \nu_{1\max} \cdot \tanh\{2\zeta(1 - t/\tau_p)\} & \tau_p/2 \leq t < \tau_p \end{cases} \quad (1)$$

In the frequency-modulated frame [62], the instantaneous frequency offset is equal to

$$\Delta\nu_0(t) = \Delta\nu_{0\max} \frac{\tan\{\kappa(1 - 2t/\tau_p)\}}{\tan(\kappa)} \quad (2)$$

where $\nu_{1\max}$ is the peak rf-amplitude, $\Delta\nu_{0\max}$ the peak rf-frequency modulation and $\zeta$ and $\kappa$ are two adjustable parameters that are used to smoothen the effective field at both pulse edges. In the absence of CSA and for on-resonance irradiation, i.e. the carrier frequency in the center of the pulse is equal to the resonance frequency, the quality factor in the first adiabatic frame is given by [59]

$$Q_1 = \frac{\nu_{1\max}^2}{\nu_R \Delta\nu_{0\max}} \frac{\pi \tan(\theta)}{\theta} \tanh^2(\zeta) \ . \quad (3)$$

Furthermore, for solids exhibiting inhomogeneous broadening, the maximum of the echo signal in a spin-echo experiment decays for increasing τ delay with a time constant $T_2'$, which is much longer than the time constant, $T_2^*$, of the free induction decay (FID). Therefore, the refocused echo can be acquired during the second τ delay, which can improve the *S*/*N* ratio by a factor of up to $\sqrt{2}$.

## III Theory

### III.1 S/N ratio with UDEFT

The analytical expression of the DEFT signal has been derived in the steady-state regime [52]. However, this expression does not allow calculating the sensitivity enhancement provided by UDEFT when only a few scans are acquired or when the initial longitudinal magnetization ($M_0$) differs from that at thermal equilibrium ($M_\infty$), e.g. when using DNP. Therefore, we derive below a more general expression (i) valid for any arbitrary initial longitudinal magnetization, and (ii) taking into account the contribution of the transient regime. The signal of UDEFT is calculated as a function of the number of scans (*NS*) for a given total experimental time, $T_{exp}$. We assume that (i) the longitudinal magnetization relaxes towards $M_\infty$ during the relaxation delay, $\tau_{RD}$, according to an exponential with $T_1$ constant-time, (ii) 2τ << $T_1$ and $\tau_{RD}$, and (iii) the longitudinal relaxation during the τ delays can be disregarded.

The total experimental time can be expressed in $T_1$ unit as $T_{exp} = AT_1 \approx NS \cdot \tau_{RD}$, which allows defining the dimensionless parameter, Ψ, as:

$$\Psi = \tau_{RD}/T_1 \approx A/NS \quad (4)$$

The efficiency, *E*, of the UDEFT sequence is defined as the fraction of the longitudinal magnetization before the first π/2-pulse, which is returned back to the z-axis after the second π-pulse. Given the above assumption, *E* can be expressed as

$$E = E_{rf} \cdot e^{(-2\tau/T_2')} \quad (5)$$

For ideal pulses, $E_{rf}$ = 1, and the total magnetization is returned back to the z-axis by the UDEFT sequence. The $e^{(-2\tau/T_2')}$ term represents the attenuation due to transverse relaxation.

Assuming that random fluctuations dominate the electronic noise, we show in the SI (Supporting Information) that the *S/N* ratio of UDEFT experiment, for which only the first FID is recorded, can be written as

$$\frac{S^{UDEFT}}{N}(NS) = K\left(\frac{e^{-\psi}}{\sqrt{NS}}\left\{M_0 - \frac{E.M_\infty(1-e^{-\psi})}{1-E.e^{-\psi}}\right\}\frac{1-(Ee^{-\psi})^{NS}}{1-E.e^{-\psi}} + \sqrt{NS}\frac{M_\infty(1-e^{-\psi})}{1-E.e^{-\psi}}\right) \quad (6)$$

where K is a constant depending on factors, such as the coil geometry, its filling factor, its temperature, its resistance, the Larmor frequency and the signal apodization [65]. As seen in Eq.6, the *S/N* ratio is the sum of two terms corresponding to the contributions of the transient (1st) and steady-state (2nd) regimes, respectively. As seen in Fig.**S1**, the transient regime only significantly contributes to the *S/N* ratio in the case of (i) short experimental time, i.e. $A = T_{exp}/T_1 \approx NS.\tau_{RD}/T_1$ is small (Fig.**S1a,b**) or (ii) hyperpolarized experiments for which $M_0$ is much larger than $M_\infty$ (Fig.**S1d**). Furthermore, when the initial magnetization is small, the FIDs acquired during the transient regime mainly contain noise and decrease the *S/N* ratio (Fig.**S1a**).

## III.2 S/N ratio with SP

For a SP experiment with a flip angle θ, we show in the SI that the *S/N* is given by

$$\frac{S^{SP}}{N}(NS) = K.\sin(\theta)\left[\frac{e^{-\psi}}{\sqrt{NS}}\left\{M_0 - \frac{\cos(\theta).M_\infty(1-e^{-\psi})}{1-\cos(\theta).e^{-\psi}}\right\}\frac{1-(\cos(\theta).e^{-\psi})^{NS}}{1-\cos(\theta).e^{-\psi}} + \sqrt{NS}\frac{M_\infty(1-e^{-\psi})}{1-\cos(\theta).e^{-\psi}}\right] \quad (7)$$

## III.3 Comparison of UDEFT and SP sensitivities

Fig.**2** shows the plot of the *S/N* of UDEFT and SP experiments versus *NS* and either *E* for UDEFT or θ for SP for A = 5 (Fig.**2a,b**) or 25 (Fig.**2c,d**). It must be noted that the S/N of UDEFT shown in Fig.**2b,d** is obtained when *only the first FID* is acquired for each scan. The acquisition of the refocused echo can increase the *S/N* ratio by a factor of up to √2, but this gain depends on *E*.

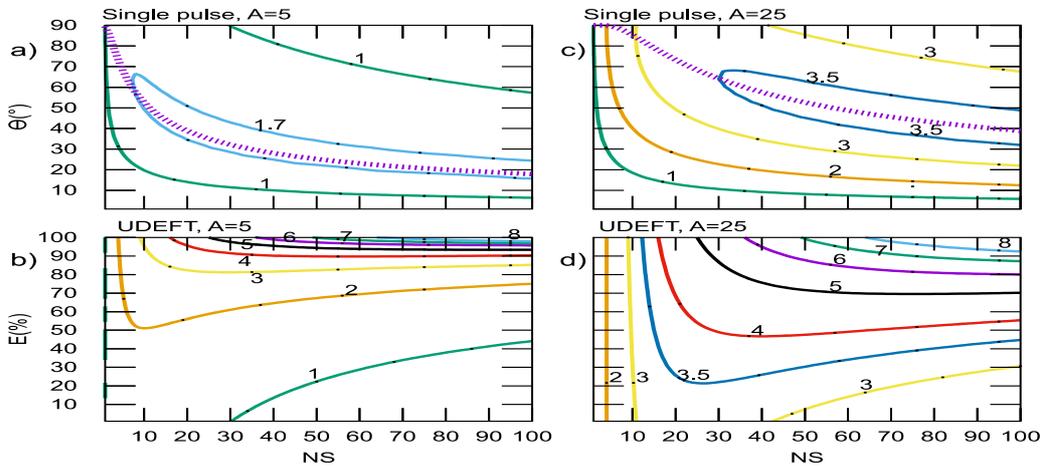

**Fig.2.** Plot of S/N of (a,c) SP and (b,d) UDEFT experiments lasting $T_{exp}$ = 5$T_1$ (a,b) or 25$T_1$ (c,d) versus *NS* and either Θ (°) for SP or *E* (%) for UDEFT. For each scan of UDEFT, only the 1st FID has been acquired. The *S/N* ratio was calculated from Eqs. 6 and 7. We assumed *K* = $M_0$ = $M_\infty$ = 1. (a,c) We also show the curve corresponding to the Ernst angle as a thick dashed purple line (Eq.8).

As expected, the *S/N* ratio increases with $T_{exp}$ and is higher in Fig.**2c** and **2d** than in **2a** and **2b**. As seen in Figs.**2a** and **2c**, the maximal *S/N* ratio for a SP experiment is achieved for a flip angle corresponding to the Ernst angle [66]:

$$\theta_E = \text{acos}[e^{-\psi}] \quad (8)$$

For small NS, the Ernst angle curves deviate from optimum conditions because of the transient contribution.

For UDEFT experiments (Figs.**2b** and **2d**), the *S/N* ratio increases with *E*. For a given *E* value, the optimal number of scans, *NS*$_{opt}$, yielding the optimal ratio, (*S/N*)$_{opt}$, increases with *T*$_{exp}$ (Table **1**). For both *T*$_{exp}$ = 5*T*$_1$ and 25*T*$_1$, the UDEFT experiments yield higher (*S/N*)$_{opt}$ than SP ones. The gain in *S/N* ratio for UDEFT with respect to SP increases with *E*. For a given *E* value, the gains are similar for both *T*$_{exp}$ = 5*T*$_1$ and 25*T*$_1$. As example, for *E* = 70%, UDEFT yields a 40% enhancement in *S/N*, which allows a two-fold reduction in T$_{exp}$.

**Table.1.** *NS*$_{opt}$ and (*S/N*)$_{opt}$ for SP and UDEFT experiments with *T*$_{exp}$ = 5*T*$_1$ or 25*T*$_1$. The *S/N* ratios were calculated from Eqs.6 and 7 with *K* = *M*$_0$ = *M*$_\infty$ = 1. For UDEFT experiment, only the 1$^{st}$ FID was acquired. For ideal pulses, *E* = 70, 90 and 95% correspond to *T*$_2$'/τ = 5.6, 19 and 39, respectively. These ratios are commonly encountered for non-protonated disordered samples [33,36].

| *T*$_{exp}$ | Scheme | *NS*$_{opt}$ | (*S/N*)$_{opt}$ | Gain in *S/N* |
|---|---|---|---|---|
| 5*T*$_1$ | SP (θ = 65-15°) | 10-100 | 1.7 | 1 |
| | UDEFT (*E* = 70 %) | 17 | 2.4 | 1.4 |
| | 90 % | 60 | 4.0 | 2.3 |
| | 95 % | 100 | 5.6 | 3.2 |
| 25*T*$_1$ | SP (θ = 65-35°) | 40-100 | 3.6 | 1 |
| | UDEFT (*E* = 70 %) | 80 | 5 | 1.4 |
| | 90 % | 250 | 8.3 | 2.3 |
| | 95 % | 500 | 11.5 | 3.2 |

### III.4 Quantitative measurements

Quantitative measurements require the longitudinal magnetization after n scans, *M*$_{R,n}$, to be close to *M*$_\infty$ for most of the scans. Hence, in the steady-state regime, we must have

$$M_{R,n} = M_{R,n-1} = mM_\infty \quad (9)$$

with *m* very close to 1. We show in Section I-3 of SI that this condition is met for UDEFT when the relaxation delay is given by

$$\tau_{RD,min} \geq T_{1,max} \cdot \ln\left(\frac{1-m.E}{1-m}\right) \quad (10)$$

with *T*$_{1,max}$ the longest *T*$_1$ value of the different sites of the sample and

$$\tau_{RD,min} \geq T_{1,max} \cdot \ln\left(\frac{1-m.\cos(\theta)}{1-m}\right) \quad (11)$$

for SP. These minimal relaxation delays and Eqs. 6 and 7 yield the maximal *S/N* ratios of

$$S/N(E,T_1) \leq K\frac{m.M_\infty}{\sqrt{\tau_{RD,min}}} = KmM_\infty \Big/ \sqrt{T_{1,max} \cdot \ln\left(\frac{1-m.E}{1-m}\right)} \quad (12)$$

for quantitative UDEFT experiments and

$$S/N(\theta,T_1) \leq K\frac{m.M_\infty.\sin(\theta)}{\sqrt{\tau_{RD,min}}} = KmM_\infty \sin(\theta) \Big/ \sqrt{T_{1,max} \cdot \ln\left(\frac{1-m.\cos(\theta)}{1-m}\right)} \quad (13)$$

for quantitative SP ones.

As seen in Fig.**3a**, quantitative SP spectra can be acquired using π/2-pulse provided $\tau_{RD,min}$ ≈ 4.5*T*$_{1,max}$. Smaller flip-angles allow the use of shorter τ$_{RD}$ delays. For example, $\tau_{RD,min}$ = 2.5*T*$_{1,max}$ for θ = 30°. Hence, to acquire quantitative SP spectra of unknown samples, it is preferable to use small θ angles. For UDEFT, $\tau_{RD,min}$ decreases inversely to *E*, and for instance, *E* ≥ 87% is required to record quantitative UDEFT spectra with $\tau_{RD,min}$ = 2.5*T*$_{1,max}$.

7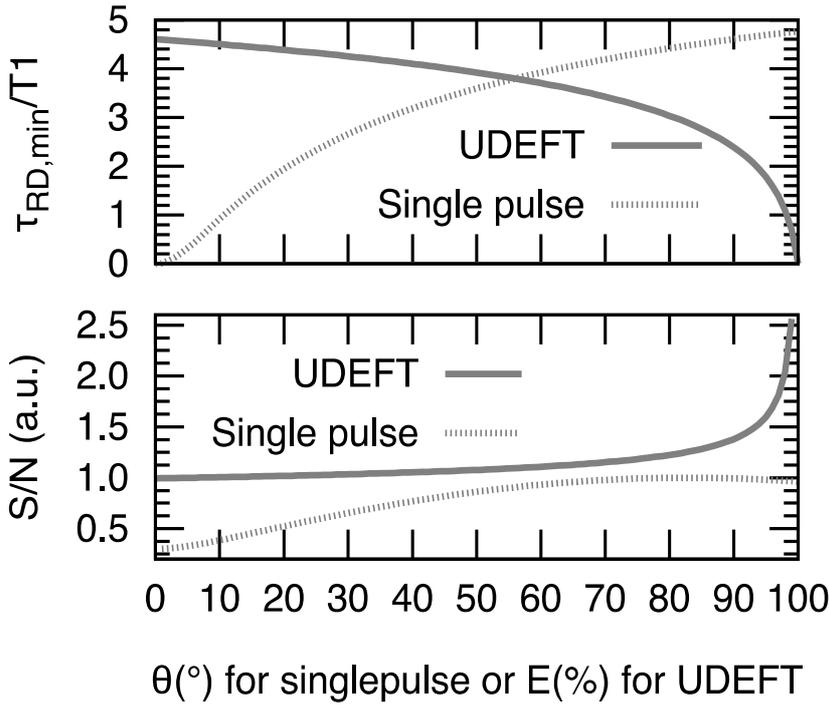

**Fig.3.** Plots of (a) $\tau_{RD,min}/T_{1,max}$ ratios (Eqs.10,11) and (b) maximal S/N ratios (Eqs.12,13) for quantitative SP and UDEFT measurements with $m = 0,99$ versus $\theta$ (°) for SP (dashed line) or $E$ (%) for UDEFT (continuous line).

Fig.**3b** shows that the maximal *S/N* ratio for quantitative SP experiments is achieved for $\theta \approx 85°$, whereas the use of $\theta = 30°$ results in a 35 % decrease. Conversely, the *S/N* of UDEFT experiment monotonously increases with *E*. The sensitivity of quantitative UDEFT experiments with $E = 87\%$ is 31 % higher than that of quantitative SP ones.

## IV Numerical simulations

### IV.1 Simulation parameters

All spin dynamics simulations were performed using the SIMPSON software [67]. The powder averaging was calculated using 1344 {$\alpha_{PR}$, $\beta_{PR}$, $\gamma_{PR}$} Euler angles describing the orientation of the principal axes of the $^{29}$Si chemical shift tensor in the rotor frame. The 168 {$\alpha_{PR}$, $\beta_{PR}$} Euler angles were selected according to the REPULSION algorithm [68], while the 8 $\gamma_{PR}$ angles were regularly stepped from 0 to 360°.

Simulations were carried out for an isolated $^{29}$Si nucleus to test the robustness of UDEFT to rf-inhomogeneity, offset and CSA using composite and adiabatic π-pulses as refocusing and inversion elements. For those simulations, the starting and detection operators were $I_z$. The static magnetic field was fixed at 9.4 T with $\nu_R$ = 4 (Tables **S1-3**, Figs.**S2**,**S3**) or 10 kHz (Figs.**4**,**S4**,**S6**).

For all simulations, τ delays were approximately equal to 2 ms and were chosen in such way that there was a multiple number of rotor periods between the centers of the π/2-pulses and that of the refocusing one. In Tables **S1** and **S2** as well as Fig.**S2**, we used ideal π/2-pulses. In Table **S1** and Fig.**S2a**, the inversion π-pulse was also ideal and in Table **S2** and Fig.**S2b**, an ideal refocusing π-pulse was used. The lengths of the pulses, which are not ideal, were calculated for a nominal rf-field $\nu_{1nom}$ = 70 kHz in Tables **S1** to **S3** and Figs.**S2**,**S3** and $\nu_{1nom}$ = 50 kHz in Figs.**4**,**S4**,**S6**.

The isotropic chemical shifts of $^{29}$Si nuclei extend from −200 to 60 ppm, which corresponds to a maximal offset of 10 kHz at $B_0$ = 9.4 T [69,70]. For Si atoms forming single covalent bonds, the CSA ranges from −60 to 90 ppm and hence, can reach 7 kHz at $B_0$ = 9.4 T [71–74]. For Si atoms forming multiple bonds, the CSA can reach −640 ppm, i.e. 50 kHz at $B_0$ = 9.4 T [6,75,76]. The rf-field produced by a solenoid coil is highly inhomogeneous [77–79], and for a 4 mm rotor it has been shown that the rf-field at the ends of the rotor, $\nu_{1edge}$, is approximately 20% of its maximal value at the center of the coil, $\nu_{1center}$ [77]. The robustness to rf-inhomogeneity (Tables **S1-3**) was investigated by varying the rf-field from 40 to 100 kHz for the refocusing and/or inversion pulses.

In Figs.**4**,**S4**, we also compared the robustness of UDEFT using as refocusing element: either a single π-pulse, denoted P180$_x$, or $59_{180}298_059_{180}$ or $58_0140_{180}344_0140_{180}58_0$ composite-π pulses, called CP$_{x1}$ and CP$_{x2}$, and as inversion element: either a single π-pulse, called P180$_z$, or $90_0240_{90}90_0$ or $90_{90}180_090_{90}$ composite π-pulses, called CP$_{z1}$ and CP$_{z2}$. We also used the adiabatic tanh/tan inversion pulse, called AP$_z$, which lasted $\tau_p$ = 50 µs and used $\Delta\nu_{0max}$ = 1.5 MHz with ζ = 10 and κ = atan (30) = 88°. It must be noted that $\tau_p$ = 25 and 100 µs gave similar results for AP$_z$ (not shown). The Simpson files used for Figs.**4**,**S4** are provided in the SI. To quantify this robustness, we have calculated the $E_{rf}$ efficiency for offset values ranging from −30 to 30 kHz and rf-fields ranging from 35 to 75 kHz, for CSA = 2 (Fig.**4**) and 20 kHz (Fig.**S4**).

## IV.2 Robustness of UDEFT to rf-field, offset and CSA

In order to improve the robustness of UDEFT to rf-field, offset and CSA, we tested seven refocusing composite π-pulses listed in Table S1, the other pulses being ideal. These pulses have a total flip angle $\theta_{tot}$ ≤ 900° and hence, for $\nu_{1nom}$ = 70 kHz, their lengths did not exceed 35 µs, a duration much shorter than the rotor period: $T_R$ = 250 µs. We tested composite π-pulses with variable or quasi-constant rotation axis designed to compensate either the rf-inhomogeneity or the offset. For constant rotation composite-π-pulses, the rotation axis remains approximately along the x axis across their effective rf-field and offset bandwidths [55,80–82]. These pulses have been shown to be better-suited than variable rotation ones for refocusing purpose in spin-echo experiments [83]. Table **S1** shows that when used as refocusing elements, the constant rotation pulses designed to invert the longitudinal magnetization with offset compensation [82], CP$_{x1}$ and CP$_{x2}$, significantly improve the robustness of UDEFT to offset with respect to P180$_x$ without deteriorating the robustness to rf-inhomogeneity. Moreover, CP$_{x2}$ better compensates for offset than CP$_{x1}$. The other tested pulses do not improve the robustness of UDEFT, or even lower it. Furthermore, Fig.**S2a** shows that CP$_{x1}$ and CP$_{x2}$ improve the robustness to CSA with respect to P180$_x$. Composite tanh/tan adiabatic pulses, including BIR-4 [61,62] and three successive adiabatic pulses with relative lengths of $\tau_P/4$, $\tau_P/2$ and $\tau_P/4$ [56], were also tested as refocusing element in UDEFT. However, they were less efficient and robust, notably to CSA, than CP$_{x1}$ and CP$_{x2}$. ==Antisymmetric composite π-pulses have been shown to act as efficient and robust refocusing elements in spin-echo experiments [84,85]. However, we have analyzed the use of these pulses with UDEFT and observed (not shown) that they are not robust to CSA owing to their very long lengths ($\theta_{tot}$ = 1620 or 2340°).==

We also tested six composite π-pulses listed in Table **S2** as inversion elements in UDEFT, the other pulses being ideal. The variable rotation pulses, CP$_{z1}$ and CP$_{z2}$, have been designed to invert the longitudinal magnetization with compensation of both rf-inhomogeneity and offset for CP$_{z1}$ and only rf-inhomogeneity for CP$_{z2}$ [54,86]. CP$_{z1}$ yields a higher robustness to offset and CSA than CP$_{z2}$ and P180$_z$ (Fig.**S2b**).

We also investigated the robustness to rf-field, offset and CSA of UDEFT built with the most robust refocusing (CP$_{x1}$ and CP$_{x2}$,) and inversion (CP$_{z1}$ and CP$_{z2}$) composite elements. The robustness of these sequences made of two composite π-pulses was compared to that of sequences using one composite and one single π-pulses, or two single π-pulses. In this case, all pulses had a finite length. Table **S3** indicates that the sequences with CP$_{x1}$-CP$_{z1}$ and CP$_{x2}$-CP$_{z1}$ pairs of composite π-pulses are the most robust to both rf-inhomogeneity and offset. The combination CP$_{x2}$-CP$_{z1}$ is more robust to offset than CP$_{x1}$-CP$_{z1}$, but this is the contrary with respect to CSA (Fig.**S3**).

Adiabatic tanh/tan pulse, AP$_z$, was also employed as inversion element. For a CSA of 2 kHz, Fig.**4** shows that AP$_z$ leads to a better robustness to rf-inhomogeneity than CP$_{z1}$ and CP$_{z2}$, and that it can be combined with $CP_{x1}$ or $CP_{x2}$ refocusing pulses in order to further improve the robustness to offset. As seen in Fig.**S4**, a larger CSA decreases the efficiency of UDEFT. Such decrease is more pronounced when $AP_z$ is combined with $CP_{x2}$ than with $CP_{x1}$, since $CP_{x1}$ is more robust to CSA than $CP_{x2}$ (Fig.**S3**). Given the typical offset and CSA values for $^{29}$Si nuclei at $B_0$ = 9.4 T ($\Delta\nu_{offset}$ ≤ 10 kHz and CSA ≤ 7 kHz for $^{29}$Si nuclei forming single covalent bonds) and the typical rf-inhomogeneity of MAS probes ($\nu_{1edge}/\nu_{1center}$ = 20%), UDEFT scheme using $CP_{x1}$ and $AP_z$ as refocusing and inversion pulses is the most robust sequence at such magnetic field.

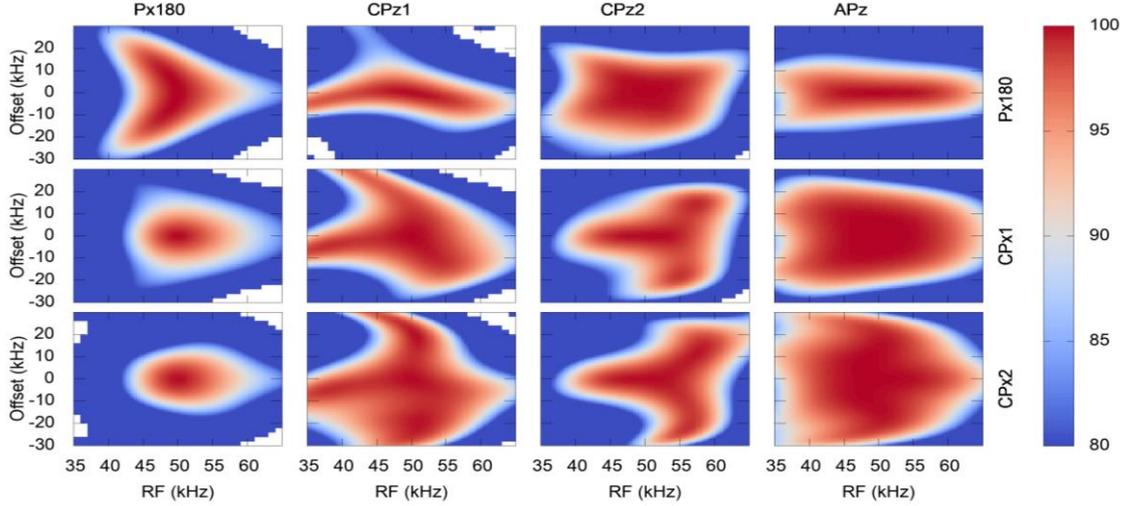

**Fig.4.** Simulated $E_{rf}$ efficiency versus rf-field and offset for UDEFT schemes using the refocusing and inversion pulses indicated on the right and the top of the figure, respectively. Simulations were performed for $^{29}$Si CSA of 2 kHz, i.e. 25 ppm at $B_0$ = 9.4 T with $\nu_{1nom}$ = 50 kHz and $\nu_R$ = 10 kHz. In these simulations, all pulses have finite lengths. The plotted $E_{rf}$ efficiency corresponds to the geometric average of two successive scans, for which the phases of refocusing and inversion pulses are incremented by 180° (caption of Fig.**1** and Section IV.3). For instance, for UDEFT scheme with $P180_x$ and $P180_z$, the 1st and 2nd scans correspond to $90_{90}$-τ-$180_0$-τ-$90_{90}$-$180_0$ and $90_{90}$-τ-$180_{180}$-τ-$90_{90}$-$180_{180}$ sequences and the plotted efficiency is equal to $E_{rf}$ = $[E_{rf}(1^{st}\,\text{scan}) \cdot E_{rf}(2^{nd}\,\text{scan})]^{1/2}$.

## IV.3 Stimulated echoes and phase cycling

Fig.**S5** displays some of the possible coherence transfer pathways during UDEFT experiments. The desired coherence transfer-pathways (Fig.**S5a**): (i) correspond to changes in coherence order of $\Delta p$ = ± 2 by each refocusing pulse, and (ii) they produce FIDs which are maximal at the beginning of the odd τ delays and at the end of the even ones. With actual refocusing pulses, changes of $p$ = ± 1, called stimulated echoes, are detected, which are maximal at the end of the odd τ delays and at the beginning of the even ones (Figs.**S5b** and **c**). The truncation of these stimulated echoes leads to undesirable oscillations around the base of the peaks. The contribution of some of these stimulated echoes to the UDEFT signal can be removed by incrementing by 180° the phase of the refocusing pulse, while the phase of the receiver remains constant. However, some of the stimulated echoes are refocused after the $\tau_{RD}$ delay, which is often much shorter than $T_1$. These echoes cannot be removed by the two-phase cycling and they produce artifacts. Stimulated echoes corresponding to $\Delta p$ = 0 by the refocusing pulse do not produce artifacts in the UDEFT spectrum (Figs.**S5d** and **e**). Furthermore, an imperfect inversion pulse can result in a residual magnetization pointing toward the –z direction during $\tau_{RD}$, which reduces the UDEFT signal.

Fig.**S6** displays the simulated $E_{rf}$ efficiency for various coherence pathways of the $CP_{x1}$-$AP_z$ sequence. These simulations show that incrementing the phase of the refocusing pulse by 180° (i) does not modify the signal intensity for the desired pathway corresponding to $\Delta p$ = ± 2, but (ii) inverts the sign of the signal corresponding to $\Delta p$ = ± 1, hence eliminating these stimulated echoes. They also show that incrementing simultaneously the phase of the refocusing π-pulse by 90° and that of the second π/2-pulse and the receiver by 180° allows removing the undesired coherence transfer pathways corresponding to $\Delta p$ = 0 during the refocusing pulse. However, as explained above, these pathways do not produce artifacts. Hence, a two-step phase cycle, in which the phase of the refocusing pulse is incremented by 180°, is sufficient. Furthermore, simulations [not shown] indicate that the $E_{rf}$ efficiency depends on the relative phase of the refocusing and inversion pulses, except in the case of adiabatic inversion pulse. Therefore, the phase of the inversion pulses is incremented simultaneously with that of the refocusing pulse, as described in the caption of Fig.**1**.

# V   Experimental results

## V.1   Experimental conditions

NMR experiments were carried with five different samples: 98% $^{29}$Si-enriched (i) amorphous silica, and (ii) borosilicate glass with $8Na_2O$-$31B_2O_3$-$61SiO_2$ molar composition prepared as described in ref. [87], or unlabeled (iii) SBA-15 mesoporous silica with a BET surface area of 650 m$^2$.g$^{-1}$ and an average pore diameter determined by BJH adsorption of 5.4 nm, (iv) mesoporous silica nanoparticles (MSNs) functionalized with 3-(N-phenylureido)propyl (PUP) groups synthesized as described in ref. [88], and (v) flame retardant material used in fire protection of steel building structures. This last material was obtained by the thermal treatment of a mixture of 92% mol of a silicone resin and 8% mol of a modifier, which is itself a mixture of polydimethylsiloxane and silica coated by a silane [89].

All experiments were acquired on a wide-bore 9.4 T Bruker NMR spectrometer equipped with an Avance-II console and a $\varnothing$ = 4 mm double resonance HX MAS probe, except in Fig.**9** ($\varnothing$ = 7 mm). The rotors were fully packed with the sample, except in Fig.**6**, and spun at $\nu_R$ = 10 kHz, except in Fig.**9** ($\nu_R$ = 5 kHz). The $T_1$ and $T_2'$ time constants were measured using saturation recovery and spin-echo experiments, respectively. $^{29}$Si 1D UDEFT spectra were recorded for all samples, CPMG ones for SBA-15 and MSNs, and SP ones for SBA-15 and flame retardant material. For UDEFT experiments, the delays between the middles of the π/2-pulses and that of the refocusing π-pulse were rotor-synchronized, i.e. equal to a multiple of the rotor period, except in Fig.**5**. Similarly, the delays between the middles of the π-pulses in CPMG experiments were also rotor-synchronized. For UDEFT scheme using AP$_z$, the adiabatic tanh/tan inversion pulse lasted $\tau_p$ = 50 μs with ζ = 10 and κ = atan (30) = 88°. No $^1$H decoupling was applied, except in Figs.**8** and **9**. The $^{29}$Si isotropic chemical shift was referenced to neat TMS. The other experimental parameters are given in the figure captions.

## V.2   Rotor synchronization of UDEFT

We first recorded the $^{29}$Si 1D UDEFT spectrum of $^{29}$Si-enriched amorphous silica, which contains approximately 90 and 10 % of Q$^4$ and Q$^3$ sites, respectively. We measured $T_1 \approx$ 55 s, $T_2' \approx$ 13 ms and a global efficiency for UDEFT sequence of $E$ = 75%. The $^2J_{Si-O-Si}$ coupling constants are typically smaller than 25 Hz [32,90], and hence the coherent signal decay produced by these *J*-couplings during a spin-echo is below 5% with τ ≈ 2 ms. This decay is taken into account in the $T_2'$ value. These *J*-couplings lead to the creation of antiphase single-quantum coherences, but their lifetime being much shorter than $\tau_{RD}$, they do not contribute to the detected signal. As seen in Fig.**5**, the intensity of UDEFT experiments is maximum when the π/2-pulses and the refocusing π-pulse are rotor-synchronized. When it is not the case, $^{29}$Si CSA and $^{29}$Si-$^{29}$Si dipolar anisotropic interactions are reintroduced and decrease the signal intensity.

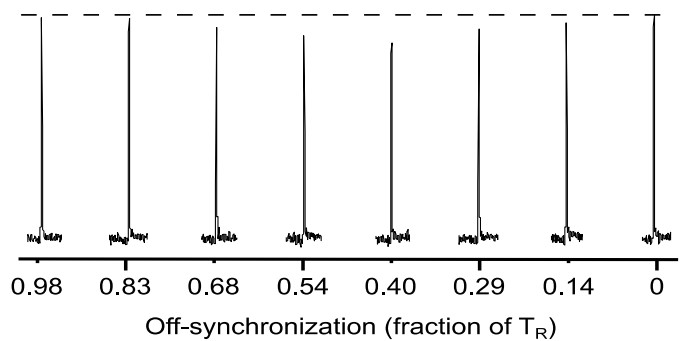

0.98   0.83   0.68   0.54   0.40   0.29   0.14   0

Off-synchronization (fraction of $T_R$)

**Fig.5.** Experimental $^{29}$Si UDEFT P180$_x$-P180$_z$ signal of $^{29}$Si-enriched amorphous silica sample versus the deviation between the delays between the 90° pulses and the refocusing pulse and a multiple of the rotor period. The horizontal dashed line facilitates the comparison of the signal intensity. $B_0$ = 9.4 T, $\nu_R$ = 10 kHz, $\nu_1$ = 75 kHz, τ = 2 ms, NS = 16, $\tau_{RD}$ = 1 s.

### V.3  Robustness to rf-field and offset of UDEFT

To test the robustness to rf-inhomogeneity, we recorded UDEFT spectra of amorphous silica using P180$_x$-P180$_z$, CP$_{x1}$-AP$_z$ and CP$_{x1}$-CP$_{z1}$ pairs of elements, versus the rf-field of the various pulses varied independently. Fig.**S7** show that for the inverting element, AP$_z$ is significantly more robust than P180$_z$ and CP$_{z1}$, whereas for the refocusing element, P180$_x$ and CP$_{x1}$ exhibit similar robustness in agreement with simulation results in Table **S1**.

In order to compare the robustness to rf-inhomogeneity and offset of UDEFT schemes using different refocusing and inversion pulses, we also recorded spectra of a borosilicate glass for various rf-field and offset values (Fig.**6**). To increase the rf-homogeneity, the sample was restricted to a slice at the center of the rotor. It must be noted that the signal which would be observed for slices at other locations where the rf-field is equal to $\nu_1$, is equal to that shown in Fig.**6b** scaled by $\nu_1/\nu_{1nom}$. Indeed, according to the reciprocity principle, the induced voltage in the coil is proportional to the rf-field [78]. In a full rotor sample, the signal would then be the integrated intensity of these curves.

This glass mostly contains Q$^3$ and Q$^4$ sites with $T_1 \approx$ 400 s and $T_2' \approx$ 10 ms. For on-resonance pulses using nominal rf-field, the global UDEFT efficiency is approximately equal to 70% for τ = 1.5 ms. UDEFT schemes with either CP$_{x1}$-AP$_z$ or CP$_{x1}$-CP$_{z1}$ exhibit similar robustness to offset but are more robust than with 180$_x$-180$_z$ (Fig.**6a**). The results shown in Fig.**6b** are consistent with the simulations of Fig.**S6** and they show that UDEFT with CP$_{x1}$-AP$_z$ is more robust to rf-field than that with CP$_{x1}$-CP$_{z1}$, which is itself more robust than with 180$_x$-180$_z$.

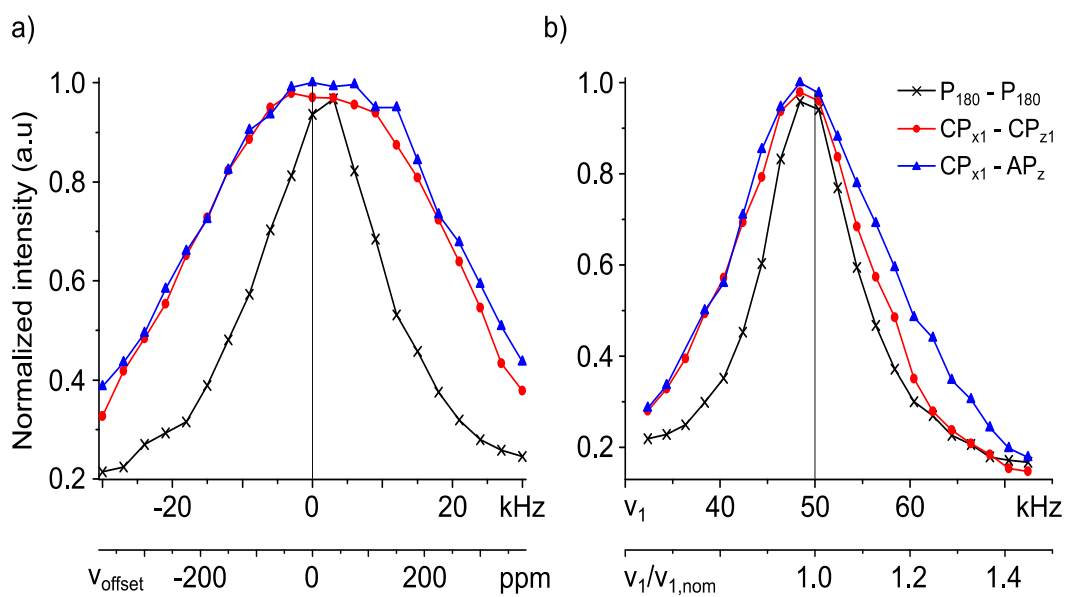

**Fig.6.** Experimental $^{29}$Si UDEFT signal of $^{29}$Si-enriched borosilicate glass versus (a) offset in kHz (top) and ppm at 9.4 T (bottom), and (b) rf-field in kHz (top) and relative value with respect to $\nu_{1nom}$ = 50 kHz (bottom) for schemes using: P180$_x$-180$_z$ (x), CP$_{x1}$-CP$_{z1}$ (●), and CP$_{x1}$-AP$_z$ (▲). $B_0$ = 9.4 T, $\nu_R$ = 10 kHz, τ = 1.5 ms, NS = 32, $\tau_{RD}$ = 5 s. To ensure identical initial magnetization, experiments started by a pre-saturation train of pulses followed with a delay of 900 s. The lengths of the single and composite pulses were calculated using $\nu_{1nom}$. For AP$_z$, $\tau_p$ = 50 μs and $\Delta\nu_{0,max}$ = 1.5 MHz. In (b), the pulses were applied on-resonance with the Q$^4$ signal.

## V.4 Comparison of UDEFT, SP and CPMG sensitivities

The sensitivities of UDEFT, SP and CPMG experiments were compared on SBA-15, which mainly contains Q$^3$ and Q$^4$ sites. The build-up curves of the $^{29}$Si longitudinal magnetization of Q$^4$ sites can be modeled as a stretched exponential with β = 0.52 and $T_1$ = 394 s (Fig.**S8**), which means that these sites exhibit a distribution of $T_1$ constant. Similarly, the decay of Q$^4$ signal in a spin-echo experiment is bi-exponential with $T'_{2f}$ = 0.34 s and $T'_{2s}$ = 1.34 s for the fast and slow components (Fig.**S9**). The distribution of $T_1$ and $T_2'$ values may stem from a faster longitudinal and transverse relaxation of Q$^4$ sites located near the surface than in the core of the silica wall.

We first compared the sensitivities of UDEFT variants using as π-pulse pairs: CP$_{x1}$-AP$_z$, CP$_{x1}$-CP$_{z1}$ and P180$_x$-P180$_z$. When using CP$_{x1}$-AP$_z$, we first optimized the number of scans (NS) to acquire the spectrum within an experimental time of $T_{exp}$ = 1 h (Fig.**7a**). The same parameters were used to acquire the spectra of Figs.**7b** and **c** for CP$_{x1}$-CP$_{z1}$, and P180$_x$-P180$_z$, respectively. The sequence using CP$_{x1}$-AP$_z$ yields higher signal than the other variants. This result is consistent with the higher robustness of this sequence to rf-inhomogeneity (Sections IV.2 and V.3). Furthermore, spectra of Figs.**7b** and **c** exhibit more intense wiggles. These truncation artifacts stem from undesirable stimulated echoes, as seen in Fig.**S10**. Indeed, P180$_z$ and CP$_{z1}$ are less robust to rf-inhomogeneity than AP$_z$, thus producing (i) an imperfect inversion in regions of the sample where the rf-field deviates from its nominal value, and hence (ii) more intense stimulated echoes. These experimental results confirm that UDEFT with CP$_{x1}$-AP$_z$ has to be preferred. Furthermore, the acquisition of the refocused FID during the second τ delay of UDEFT experiment allows enhancing the sensitivity by $\sqrt{2}$, as seen in Fig.**7e**. This gain corresponds to the theoretical limit since this sample features long $T_2'$ values and the losses during 2τ are limited. We then compared these spectra with those obtained with SP and CPMG. For SP experiment, the pulse length and the number of scans were optimized to maximize the sensitivity, which provided the spectrum shown in Fig.**7d**. Its sensitivity is approximately 9-fold lower than that of the best UDEFT version (Fig.**7e**). Conversely, for that sample featuring very long $T_2'$ values, the sensitivity of CPMG experiment is 70% higher than that of UDEFT (compare Figs.**7e** and **f**).

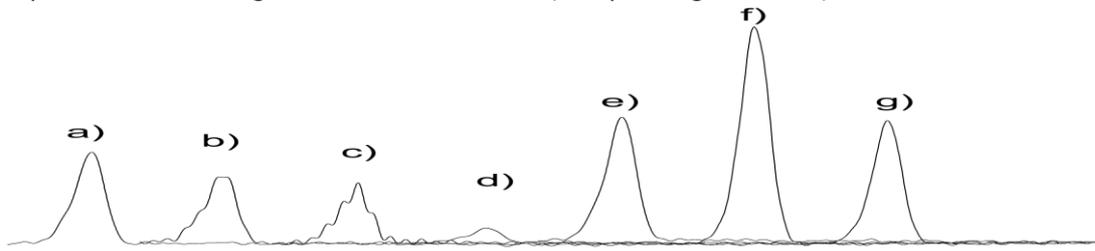

**Fig.7.** $^{29}$Si MAS spectra of SBA-15 acquired using (a-c, e) UDEFT, (d) SP and (f,g) CPMG experiments. $B_0$ = 9.4 T, $\nu_R$ = 10 kHz, τ = 2.7 ms, $T_{exp}$ = 1 h. To ensure identical initial magnetization, experiments started by a pre-saturation train of pulses followed with a 1 h delay. **UDEFT:** NS = 2048, $\tau_{RD}$ = 1.75 s, with (a-c) only the 1$^{st}$ or (e) also the 2$^{nd}$ FID. (a,e) CP$_{x1}$-AP$_z$, (b) CP$_{x1}$-CP$_{z1}$, (c) P180$_x$-P180$_z$. AP$_z$: $\tau_p$ = 50 μs, $\nu_{1,max}$ = 52 kHz, $\Delta\nu_{0,max}$ = 2.5 MHz. The rf-field of other pulses was 52 kHz. **SP:** $\tau_p$ = 2.1 μs (flip-angle = 40°), $\nu_1$ = 46 kHz, NS = 128, $\tau_{RD}$ = 28 s. **CPMG:** $\nu_1$ = 52 kHz, NS = 4, n = 4096 (f) or 256 (g). The intensities were carefully normalized to the same S/N ratio as in spectra (a-c) with NS = 2048: they were multiplied by $1/\sqrt{2}$ (e,f), 4 (d) and $2\sqrt{2}$ (g).

We also compared the sensitivity of UDEFT and CPMG experiments for the sample of MSNs functionalized with PUP groups. As seen in Fig.**8**, the $^{29}$Si spectrum of that sample exhibits three resolved resonance ascribed to T, Q$^3$ and Q$^4$ sites [91,92]. The concentration of protons in the pores is much higher for that sample than for SBA-15, and hence $^1$H decoupling must be applied during the τ delays of UDEFT for resolution purpose and the full CPMG sequence to detect the refocused echoes. For UDEFT, the spin-echoes lasting 2τ are interleaved with $\tau_{RD}$ delays, during which no $^1$H decoupling is applied and the decoupling periods thus remain very short. For CPMG, the number of echoes acquired for each scan is limited by the power-handling specifications of the probe. For functionalized MSNs, only ten echoes could be acquired with CPMG so that the decoupling period does not exceed 50 ms. The comparison of the UDEFT and CPMG spectra shown in Fig.**8a** indicates that UDEFT, for which only the 1st FID is acquired, is approximately 50 and 20 % more sensitive than CPMG for T and Q sites, respectively. UDEFT yields larger enhancement for the T sites than for the Q ones since the former are subject to larger $^1$H-$^{29}$Si dipolar couplings and their CPMG signal decays more rapidly. Furthermore, a 25% additional gain in sensitivity can be obtained by acquiring the 2nd FID with UDEFT. Hence, for these functionalized MSNs, UDEFT with acquisition of the two FIDs is approximately 80 and 50% more sensitive than CPMG for the detection of T and Q sites, respectively.

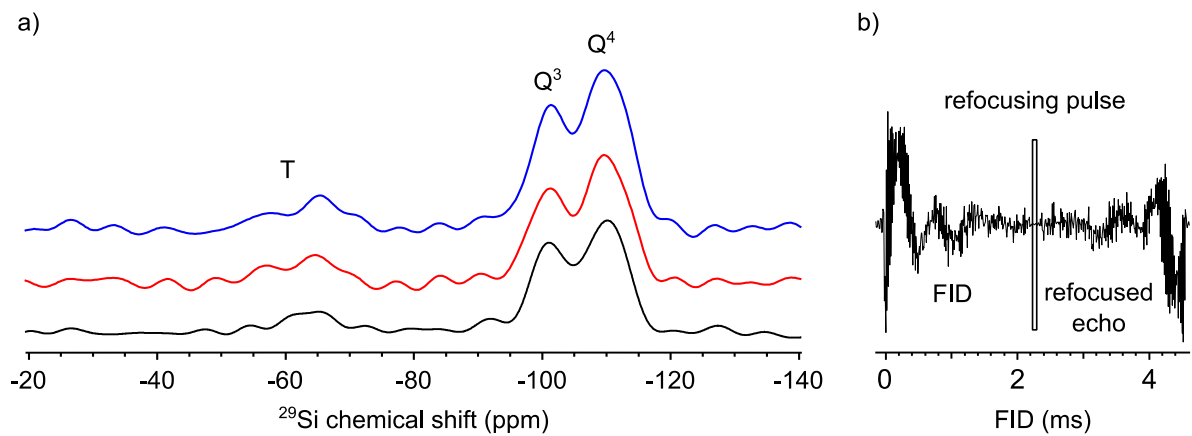

**Fig.8.** (a) $^{29}$Si MAS spectra of MSNs functionalized with PUP acquired with CPMG (bottom-black) or UDEFT CP$_{x1}$-AP$_z$ with either only the 1st (middle-red) or also the 2nd-FID (top-blue). The signal intensities are scaled to have the same noise. SPINAL-64 $^1$H decoupling with an rf-field of 80 kHz was applied during the delays of τ = 2.2 ms. $B_0$ = 9.4 T, $\nu_R$ = 10 kHz, $T_{exp}$ = 7 h, $\nu_1$ = 92 kHz (except AP$_z$). **CPMG:** NS = 262, $\tau_{RD}$ = 96 s, 10 echoes (limited by 50 ms acquisition to save the probe from $^1$H decoupling). **UDEFT:** NS = 2096, $\tau_{RD}$ = 12 s. AP$_z$: $\tau_p$ = 50 μs, $\nu_{1max}$ = 58 kHz, $\Delta\nu_{0max}$ = 4 MHz. (b) FIDs of UDEFT experiments.

## V.5 Quantitative spectra

We also compared the quantification of $^{29}$Si signals for the flame retardant material. The $^{29}$Si spectrum of this sample exhibits three signals at −22, −70 and −78 ppm, which are attributed to D sites of polydimethylsiloxane and T$^2$ and T$^3$ sites of silicone polymer, respectively. A very weak Q$^4$ site at ca. −110 ppm is observed for long acquisition time. The $T_1$ times of D and T sites ranges from 40 to 70 s. The $^{29}$Si polydimethylsiloxane D site signal at −22 ppm is narrow with a linewidth below 30 Hz, but it was broadened to 100 Hz by an exponential multiplication in Fig.**9**. The $^1$H signals of those molecules are also very narrow, which indicates their high mobility. These fast and isotropic motions of polydimethylsiloxane chains are confirmed by the lack of D signal in $^1$H → $^{29}$Si CPMAS experiments, which indicates vanishing $^1$H-$^{29}$Si dipolar couplings. On the contrary, the T sites exhibit broad signals, which are visible in $^1$H → $^{29}$Si CPMAS spectra. Hence, the silicone polymers are rigid.

Fig.**9** compares the $^{29}$Si MAS spectra of this material acquired with SP experiments using $\tau_{RD}$ = 180 s and with UDEFT using $\tau_{RD}$ from 12.5 to 150 s, with the same number of scans. The intensity of UDEFT signals reaches that of SP for $\tau_{RD}$ = 50 s for the D site and $\tau_{RD}$ = 75 s for the T site. This result confirms the higher sensitivity of UDEFT experiment with respect to SP. The employed ⌀ = 7 mm probe produces highly inhomogeneous rf-field and UDEFT would yield better sensitivity gain using 4 mm probe. Furthermore, with UDEFT the intensity of the T site keeps increasing for $\tau_{RD} \geq$ 75 s and exceeds that of SP spectrum. This result indicates that UDEFT spectra acquired within a shorter experimental time than the SP ones yield better quantification of the various sites. A quantitative analysis of the proportions and S/Ns of the $^{29}$Si MAS spectra of flame retardant material shown in Fig.**9**. is given in Table S4.

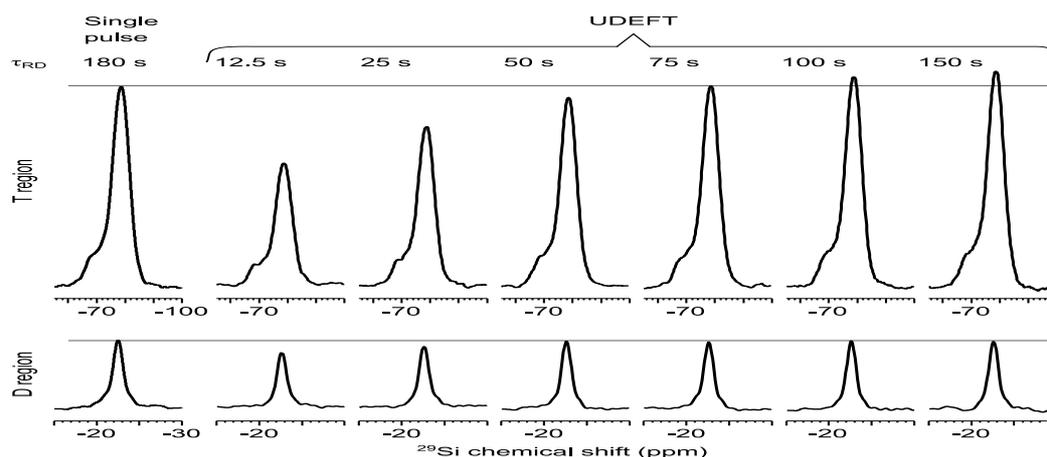

**Fig.9.** $^{29}$Si MAS signals of D (bottom) and T (top) species of the flame retardant material acquired with SP (left) or UDEFT CP$_{x1}$-AP$_z$ (right) techniques. $B_0$ = 9.4 T, $\nu_R$ = 5 kHz, NS = 80, $\nu_1 = \nu_{1,max}$ = 45 kHz for all pulses of SP and UDEFT experiments. SPINAL-64 $^1$H dipolar decoupling with rf-field of 40 kHz was applied during the 10 ms of FID (SP) or τ delays (UDEFT). **SP:** the FID was acquired after a π/2-pulse. **UDEFT**: only the 1$^{st}$ FID was used to be quantitative. AP$_z$: $\tau_p$ = 50 μs, $\Delta\nu_{0max}$ = 3 MHz. $\tau_{RD}$ delays are indicated above the spectra. The signal was multiplied by a decaying exponential function (Lorentzian broadening of 80 Hz) to remove residual truncation artifacts.

## VI. Conclusions

We have demonstrated herein the possibility to acquire $^{29}$Si MAS NMR spectra of solids using UDEFT experiment. We have shown that the use of 59$_{180}$298$_0$59$_{180}$ refocusing composite π-pulse and adiabatic inversion pulse using tanh/tan modulation improves the efficiency of this sequence and its robustness to rf-inhomogeneity, offset and CSA. These pulses combined with the phase cycling limit the artifacts produced by stimulated echoes. We have theoretically and experimentally demonstrated the gain in sensitivity provided by UDEFT with respect to SP experiments for disordered samples with $T_2^* < T_2'$. The main limitation of UDEFT, and also CPMG, may be when the sample is very well crystallized with a small number of narrow resonances. Indeed, in this case the previous condition, $T_2^* < T_2'$, may not be met, which leads to very long delays and possible truncation effects. In the case of protonated samples, UDEFT experiment can also be more sensitive than CPMG since the power handling specification of the probe limits the maximal length of the $^1$H decoupling and the number of echoes acquired during each scan of CPMG sequence. Furthermore, UDEFT sequence yields a better quantification of the NMR signals than SP, and a fortiori CPMG, while offering a higher sensitivity than quantitative SP experiments.

Using large rotor diameters (e.g. ⌀ = 4 or 7 mm) for UDEFT is useful for sensitivity reasons, especially in the case of unlabeled samples with very long relaxation times of several hundreds or thousands of seconds. However, for less demanding samples other options could be used, such as ⌀ = 3.2 mm rotors.

**Acknowledgments.** Chevreul Institute (FR 2638), Ministère de l'Enseignement Supérieur, de la Recherche et de l'Innovation, Region Hauts-de-France and FEDER are acknowledged for supporting and funding partially this work. Financial support from the IR-RMN-THC FR-3050 CNRS for conducting the research is gratefully acknowledged. Authors also thank contracts ANR-17-ERC2-0022 (EOS) and ANR-18-CE08-0015-01 (ThinGlass). This project has received funding from the European Union's Horizon 2020 research and innovation program under grant agreement No 731019 (EUSMI). OL acknowledge financial support from Institut Universitaire de France (IUF). Authors thank Jean-Sebastien Girardon from UCCS at the Lille-University for giving them the SBA-15 sample and Igor Slowing and Marek Pruski from Ames Laboratory (USA) for providing them the silica sample functionalized with 3-(N -phenylureido)propyl groups

In the SI, we provide: Figs.S1-S10, Tables S1-S3 and the SIMPSON input files for Figs.4 and S4.
The pulse sequences for various Bruker consoles are provided in a separate archive.

# Improved sensitivity and quantification for $^{29}$Si NMR experiments on solids using UDEFT (Uniform Driven Equilibrium Fourier Transform)

Nghia Tuan Duong,[1,#] Julien Trébosc,[1,2*] Olivier Lafon,[1,3] Jean-Paul Amoureux[1,4*]

# Supporting Information

[1] Univ. Lille, Centrale Lille, ENSCL, Univ. Artois, CNRS-8181, UCCS – Unit of Catalysis and Chemistry of Solids, F-59000 Lille, France.
[2] Univ. Lille, CNRS-FR2638, Fédération Chevreul, F-59000 Lille, France.
[3] Institut Universitaire de France, 1 rue Descartes, F-75231 Paris, France.
[4] Bruker Biospin, 34 rue de l'industrie, F-67166 Wissembourg, France.
[#] Present address: RIKEN-JEOL Collaboration Center, Yokohama, Kanagawa 230-0045, Japan.

## I. Derivation of Eqs. 3 and 9

For both UDEFT and SP sequences, the longitudinal magnetization after $n \leq NS$ scans at the beginning of the $\tau_{RD}$ delay is denoted $M_n$. It relaxes during $\tau_{RD}$ to the $M_{R,n}$ longitudinal magnetization given by:

$$M_{R,n} = M_n e^{-\Psi} + M_\infty (1 - e^{-\Psi}) \tag{S1}$$

where $\Psi$ is defined in Eq. 4.

### I-1. UDEFT sequence
#### I-1-1. Magnetization

For UDEFT sequence, the longitudinal magnetization after $n+1$ scans is equal to:

$$M_{n+1} = E \cdot M_{R,n} \tag{S2}$$

where the UDEFT efficiency $E$ is given by Eq. 5. From Eqs. S1 and S2, we can deduce the following recurrence relation:

$$M_{n+1} = E\{M_n e^{-\Psi} + M_\infty(1 - e^{-\Psi})\} = B + CM_n \tag{S4}$$

with

$$B = E \cdot M_\infty(1 - e^{-\Psi}) \text{ and } C = E e^{-\Psi} \tag{S5}$$

Eq. S4 is a first-order linear difference equation and $M_n$ can be expressed as

$$M_n = \frac{B}{1-C} + \left\{M_0 - \frac{B}{1-C}\right\} C^n \tag{S6}$$

#### I-1-2. Signal to noise ratio

The signal of the $n^{th}$ scan, $s_n$, is proportional to the $M_{R,n}$ magnetization

$$s_n = D M_{R,n} \tag{S7}$$

where $D$ is a constant subsuming several factors, such as coil geometry, filling factor, Larmor frequency and apodization [1]. Using Eq. S2, the $s_n$ signal can also be written

$$s_n = \frac{D}{E} M_n \tag{S8}$$

The total signal after the $NS$ scans, $S_{NS}$, of the UDEFT experiment can be expressed as

$$S_{NS} = \sum_{n=1}^{NS} s_n = \frac{D}{E}\sum_{n=1}^{NS} M_n \qquad (S9)$$

using Eq. S8. By substituting Eq. S6 into Eq. S9, $S_{NS}$ appears as the sum of an arithmetic series and a geometric one with a common ratio $C$ and can thus be written as

$$S_{NS} = \frac{D}{E}\left\{NS\frac{B}{1-C} + \left\{M_0 - \frac{B}{1-C}\right\}\frac{1-C^{NS}}{1-C}\right\} \qquad (S10)$$

By substituting $B$ and $C$ constants by their expressions given in Eq. S5, we obtain

$$S^{UDEFT}(NS) = D\left(e^{-\psi}\left\{M_0 - \frac{EM_\infty(1-e^{-\psi})}{1-Ee^{-\psi}}\right\}\frac{1-(Ee^{-\psi})^{NS}}{1-Ee^{-\psi}} + NS\frac{M_\infty(1-e^{-\psi})}{1-Ee^{-\psi}}\right) \qquad (S11)$$

This equation is the sum of two contributions. The last term corresponds to the steady-state regime and provides the same signal every scan. The first term corresponds to transient signals and rapidly converges to a constant value when $NS \gg 1$, $(Ee^{-\psi})^{NS} \to 0$, and thus the accumulated signal becomes:

$$S^{UDEFT}(NS \gg 1) = D\left(e^{-\psi}\frac{M_0 - EM_\infty + Ee^{-\psi}(M_\infty - M_0)}{(1-Ee^{-\psi})^2} + NS\frac{M_\infty(1-e^{-\psi})}{1-Ee^{-\psi}}\right) \qquad (S12)$$

Assuming the noise in UDEFT experiment is mainly random, its root-mean-square (rms) amplitude is proportional to $\sqrt{NS}$:

$$N(NS) = N(1)\sqrt{NS} \qquad (S13)$$

where $N(1)$ is the rms amplitude of the noise for a single scan. This amplitude depends on the temperature of the coil, its resistance and the bandwidth of the receiver [1]. The $S/N$ can be calculated by dividing Eq. S11 by Eq. S13, which yields Eq. 6 with $K = D/N(1)$.

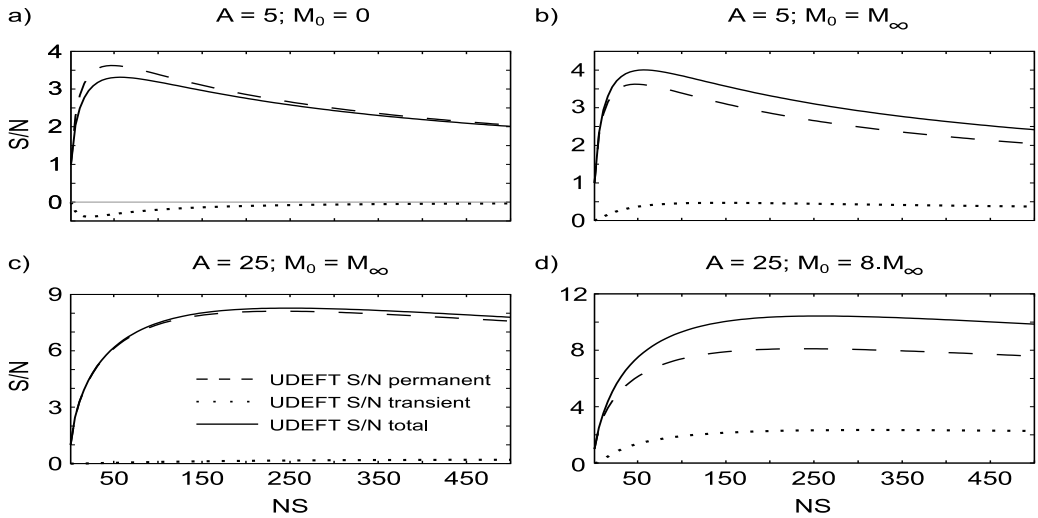

**Fig.S1.** Plot of total S/N ratio (continuous line) of UDEFT experiment as well as the contributions to S/N of transient regime (dotted line) and steady-state one (dashed line) as function of NS for E = 90 % and A = (a,b) 5 or (c,d) 25. The initial magnetization is $M_0$ = 0 (a), $M_\infty$, (b,c) or $8M_\infty$. (d). The total S/N ratio, the transient contribution and the steady-state one were calculated using Eq. 6, the first term and the second one, respectively, with K = $M_\infty$ = 1.

## I-2. SP sequence

### I-2-1. Magnetization

After the $n^{th}$ θ pulse of SP experiments, the longitudinal magnetization is given by

$$M_{n+1} = \cos(\theta)M_{R,n} \qquad (S14)$$

By combining Eqs. S1 and S14, we obtain

$$M_{n+1} = \cos(\theta)\{M_n e^{-\Psi} + M_\infty(1 - e^{-\Psi})\} = B' + C'M_n \qquad (S15)$$

with

$$B' = \cos(\theta)M_\infty(1 - e^{-\Psi}) \text{ and } C' = \cos(\theta)e^{-\Psi} \tag{S16}$$

Eq. S15 is similar to Eq. S4 and hence, $M_n$ can be expressed as

$$M_n = \frac{B'}{1-C'} + \left\{M_0 - \frac{B'}{1-C'}\right\}C'^n. \tag{S17}$$

### I-2-2. Signal to noise ratio

The signal of the $n^{\text{th}}$ scan, $s_n$, is given by

$$s_n = D\sin(\theta)M_{R,n} \tag{S18}$$

By substituting Eq. S1 into Eq. S18, we find that the total signal $S_{NS} = \sum_{n=1}^{NS} s_n$ is equal to

$$S_{NS} = D\sin(\theta)\left[e^{-\Psi}\left\{M_0 - \frac{\cos(\theta)M_\infty(1-e^{-\Psi})}{1-\cos(\theta)\cdot e^{-\Psi}}\right\}\frac{1-(\cos(\theta)e^{-\Psi})^{NS}}{1-\cos(\theta)e^{-\Psi}} + NS\frac{M_\infty(1-e^{-\Psi})}{1-\cos(\theta)e^{-\Psi}}\right] \tag{S19}$$

Eq. S13 is still valid for SP experiment and the S/N ratio is given by Eq. 7.

### I-3. Quantitative measurements

For UDEFT, by substituting Eqs. S1 and S2 into Eq. 9, we obtain

$$m = m.E.e^{-\Psi} + (1-e^{-\Psi}) \tag{S20}$$

which can be recast into Eq. 10. For SP, an equation similar to Eq. S20 with $E$ replaced by $\cos(\theta)$ can be obtained by substituting Eqs. S1 and S14 into Eq. 6 and can be recast into Eq. 11.

## II. Numerical simulations

### II-1. Optimization of composite π-pulses

#### II-1-1. Robustness to rf-field and offset for the refocusing π-pulse

| Refocusing π-pulse | Ref. | $\theta_{tot}$ /° | Rotation | Compensation | $\frac{\Delta\nu_1}{\nu_{1nom}}$ (a) | $\frac{\Delta\nu_0}{\nu_{1nom}}$ (b) |
|---|---|---|---|---|---|---|
| $180_0$ (P180$_x$) | | 180 | variable | | 0.86 | 0.46 |
| $59_{180}298_059_{180}$ (CP$_{x1}$) | [2] | 416 | constant | $\nu_0$ | 0.86 | **0.86** |
| $58_0140_{180}344_0140_{180}58_0$ (CP$_{x2}$) | [2] | 740 | constant | $\nu_0$ | 0.86 | **1.11** |
| $180_{120}180_{240}180_{120}$ (CP$_{x3}$) | [3] | 540 | constant | $\nu_1$ | *0.76* | *0.31* |
| $90_{90}180_090_{90}$ (CP$_{x4}$) | [4] | 360 | variable | $\nu_1$ | *0.29* | **0.51** |
| $90_0360_{120}90_0$ (CP$_{x5}$) | [5] | 540 | variable | $\nu_1$ | *0.14* | **0.80** |
| $180_{104.5}360_{313.4}180_{104.5}180_0$ (CP$_{x6}$) | [6] | 900 | variable | $\nu_1$ | 0.86 | **0.60** |
| $90_0255_{180}315_0$ (CP$_{x7}$) | [7] | 660 | variable | $\nu_0$ | 0.86 | *0.23* |

**Table S1**. Robustness to rf-field and offset of UDEFT using the refocusing pulses listed in the first column. The ranges of rf-fields and offsets yielding $E_{rf} \geq 90\%$ were determined using spin-dynamics simulations with $\nu_R = 4$ kHz and $\nu_{1nom} = 70$ kHz. In these simulations, all pulses of UDEFT are ideal, except for the refocusing π-pulse. Bandwidths which are extended with respect to $180_0$ are in bold type, whereas those which are smaller are in italics. **(a)** Range of rf-fields for which $E_{rf} \geq 90\%$, normalized with respect to $\nu_{1nom}$. The $\nu_1$ values yielding $E_{rf} = 90\%$ are symmetrical with respect to $\nu_{1nom}$. **(b)** Range of offsets for which $E_{rf} \geq 90\%$ normalized with respect to resonance frequency. The offset values yielding $E_{rf} = 90\%$ are symmetrical with respect to resonance frequency, except for $CP_{x6}$, for which a slight asymmetry is observed.

## II-1-2. Robustness to rf-field and offset for the inversion π-pulses

| Inversion π-pulse | Ref. | $\theta_{tot}$ /° | Rotation | Compensation | $\frac{\Delta\nu_1}{\nu_{1nom}}$ (a) | $\frac{\Delta\nu_0}{\nu_{1nom}}$ (b) |
|---|---|---|---|---|---|---|
| $180_0$ ($P180_z$) | | 180 | variable | | 0.29 | 0.46 |
| $90_0 240_{90} 90_0$ ($CP_{z1}$) | [8] | 420 | variable | $\nu_0$ and $\nu_1$ | **0.61** | **1.29** |
| $90_{90} 180_0 90_{90}$ ($CP_{z2}$) | [4] | 360 | variable | $\nu_1$ | **0.63** | **0.51** |
| $180_{120} 180_{240} 180_{120}$ ($CP_{z3}$) | [3] | 540 | variable | $\nu_1$ | **0.83** | *0.23* |
| $58_0 140_{180} 344_0 140_{180} 58_0$ ($CP_{z4}$) | [2] | 740 | constant | $\nu_0$ | 0.29 | **1.00** |
| $59_{180} 298_0 59_{180}$ ($CP_{z5}$) | [2] | 416 | constant | $\nu_0$ | 0.29 | **0.74** |
| $90_0 360_{120} 90_0$ ($CP_{z6}$) | [5] | 540 | variable | $\nu_1$ | **0.83** | *0.40* |

**Table S2**. Robustness to rf-field and offset of UDEFT using the inversion pulses listed in the first column. The ranges of rf-fields and offsets yielding $E_{rf} \geq 90\%$ were determined using spin dynamics simulations with $\nu_R = 4$ kHz and $\nu_{1nom} = 70$ kHz. In these simulations, all pulses are ideal, except for the inversion π-pulse. Bandwidths which are extended with respect to $180_0$ are in bold type, whereas those which are smaller are in italics. **(a)** Range of rf-fields for which $E_{rf} \geq 90\%$, normalized with respect to $\nu_{1nom}$. The $\nu_1$ values yielding $E_{rf} = 90\%$ are symmetrical with respect to $\nu_{1nom}$, except for $CP_{z1}$, for which an asymmetry is observed. **(b)** Range of offset values for which $E_{rf} \geq 90\%$ normalized with respect to resonance frequency.

## II-1-3. Robustness to rf-field and offset for the pairs of composite π-pulses

| Combination of pulses | $E_{rf} \geq 90\%$ | | $E_{rf} \geq 98\%$ | |
|---|---|---|---|---|
| | $\frac{\Delta\nu_1}{\nu_{1nom}}$ (a) | $\frac{\Delta\nu_0}{\nu_{1nom}}$ (b) | $\frac{\Delta\nu_1}{\nu_{1nom}}$ (a) | $\frac{\Delta\nu_0}{\nu_{1nom}}$ (b) |
| $P180_x$-$P180_z$ | 0.29 | 0.77 | 0.14 | 0.49 |
| $P180_x$-$CP_{z1}$ | **0.61** | *0.40* | **0.44** | *0.16* |
| $P180_x$-$CP_{z2}$ | **0.63** | **0.80** | **0.43** | **0.50** |
| $CP_{x1}$-$P180_z$ | 0.29 | *0.57* | 0.14 | *0.21* |
| $CP_{x1}$-$CP_{z1}$ | **0.61** | **0.93** | **0.44** | **0.56** |
| $CP_{x1}$-$CP_{z2}$ | **0.63** | **0.83** | **0.43** | *0.21* |
| $CP_{x2}$-$P180_z$ | 0.29 | *0.47* | 0.14 | *0.24* |
| $CP_{x2}$-$CP_{z1}$ | **0.61** | **1.20** | **0.44** | **1.10** |
| $CP_{x2}$-$CP_{z2}$ | **0.63** | *0.51* | **0.43** | *0.23* |

**Table S3**. Robustness to rf-field and offset of UDEFT using the combination of refocusing and inversion pulses listed in the first column. The ranges of rf-fields and offsets yielding $E_{rf} \geq 90$ or 98 % were determined using spin dynamics simulations with $\nu_R = 4$ kHz and $\nu_{1nom} = 70$ kHz. In these simulations, all pulses have a finite length. Bandwidths which are extended with respect to $180_0$ are in bold type. Bandwidths which are contracted with respect to $180^0$ are in italics. **(a)** Range of rf fields for which $E_{rf} \geq 90\%$ (2$^{nd}$ column) and 98% (4$^{th}$ column) normalized with respect to $\nu_{1nom}$. The $\nu_1$ values yielding $E_{rf}$ = 90 and 98 % are symmetrical with respect to $\nu_{1nom}$. **(b)** Range of offsets for which $E_{rf} \geq 90\%$ (3$^{rd}$ column) and 98% (5$^{th}$ column) normalized with respect to resonance frequency.

## II-1-4. Robustness to CSA for the composite π-pulses

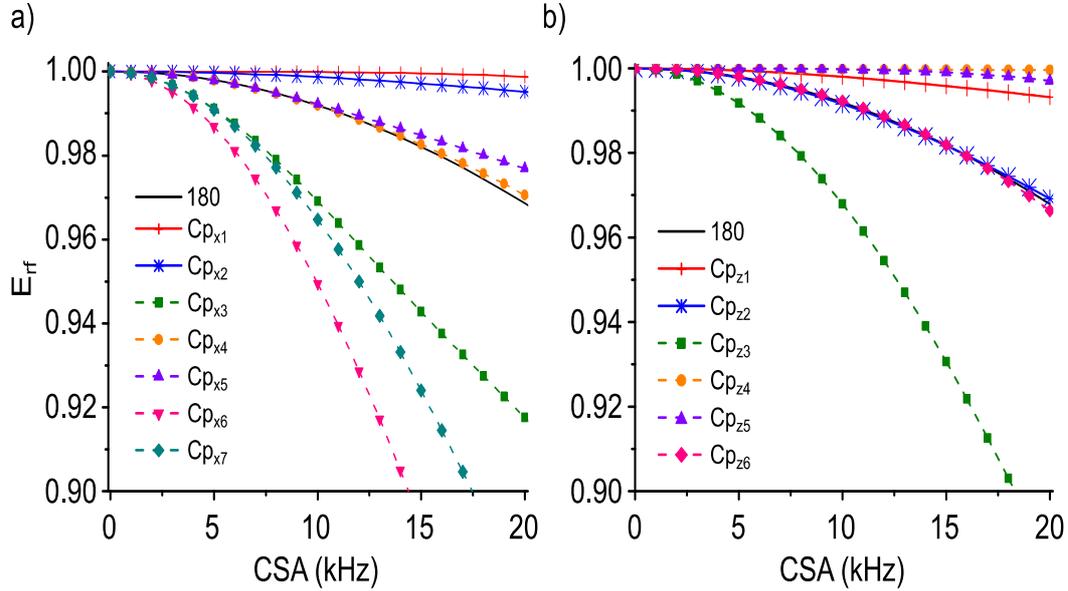

**Fig.S2.** Robustness to CSA of UDEFT using the (a) refocusing and (b) inversion composite π-pulses listed in Tables **S1** and **S2**. All pulses were applied on resonance using $\nu_{1nom}$ = 70 kHz with $\nu_R$ = 4 kHz. In these simulations, all pulses were ideal, except the (a) refocusing or (b) inversion composite π-pulses.

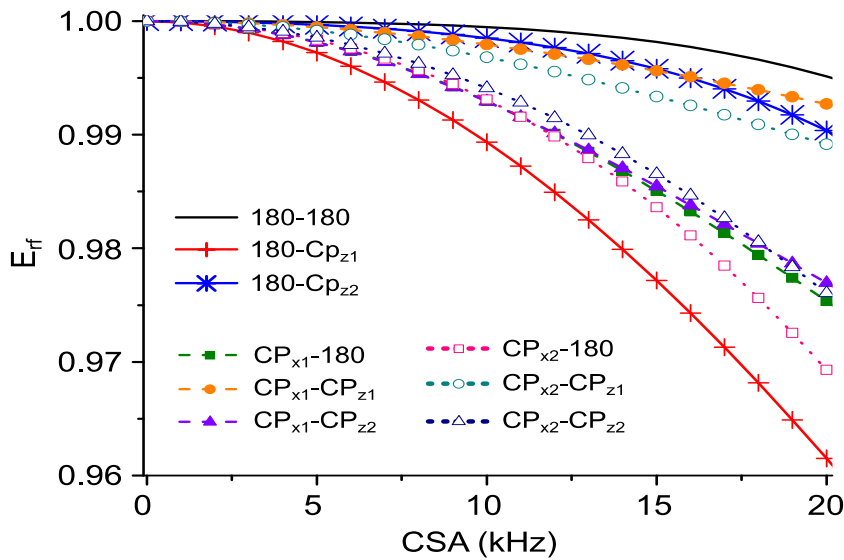

**Fig.S3.** Robustness to CSA of UDEFT using different pairs of refocusing-inversion composite π-pulses listed in Table **S3**. All pulses were applied on resonance using $\nu_{1nom}$ = 70 kHz with $\nu_R$ = 4 kHz. In these simulations, all pulses have a finite length.

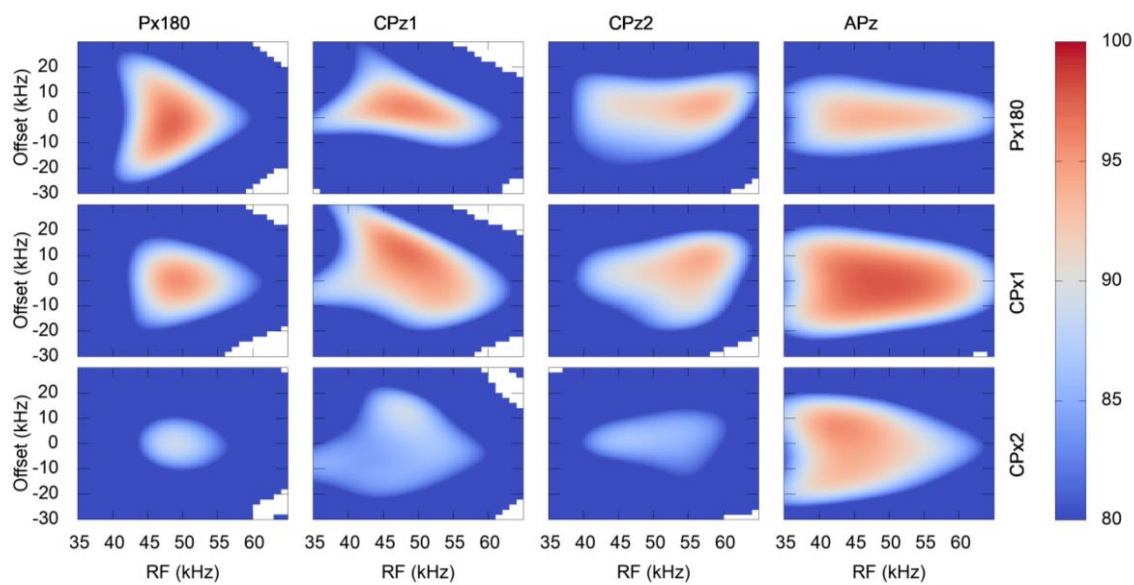

**Fig.S4.** Simulated $E_{rf}$ efficiency versus rf-field and offset for UDEFT schemes using the refocusing and inversion pulses indicated on the right and the top of the figure, respectively. Simulations were performed for $^{29}$Si CSA of 20 kHz, i.e. 250 ppm at $B_0$ = 9.4 T with $\nu_{1nom}$ = 50 kHz and $\nu_R$ = 10 kHz. In these simulations, all pulses have a finite length. The plotted $E_{rf}$ efficiency corresponds to the geometric average of $E_{rf}$ efficiencies of two successive scans, for which the phases of both refocusing and inversion pulses are incremented by 180° (see caption of Fig.**4**).

## II-2. Stimulated echoes and phase cycling

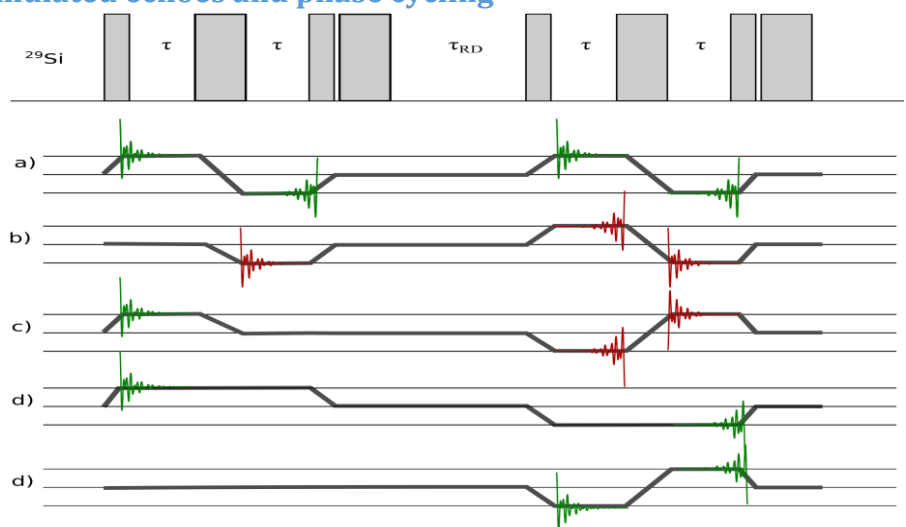

**Fig.S5.** (a-e) Examples of coherence transfer pathways leading to observable signals during the τ delays. The pulses and delays corresponding to two successive scans of UDEFT are displayed on the top. The FIDs contributing to the wanted signals are displayed in green, whereas those producing artifacts are displayed in red. (a) One of the desired coherence transfer pathways. (b-c) Coherence transfer pathways producing artifacts. (b) The contribution of the 1st FID to the detected signal is eliminated by the two-step phase cycling described in the caption of Fig.**1**. However, the 2nd and 3rd FIDs in (b) and (c) correspond to Δp = ±2 for the refocusing pulse as for the desired coherence transfer pathway and hence, cannot be removed by the phase cycle. (d,e) Coherence transfer pathways resulting from pulse imperfections, which do not produce artifacts in the UDEFT spectrum. Symmetric coherence transfer pathways produce identical signals.

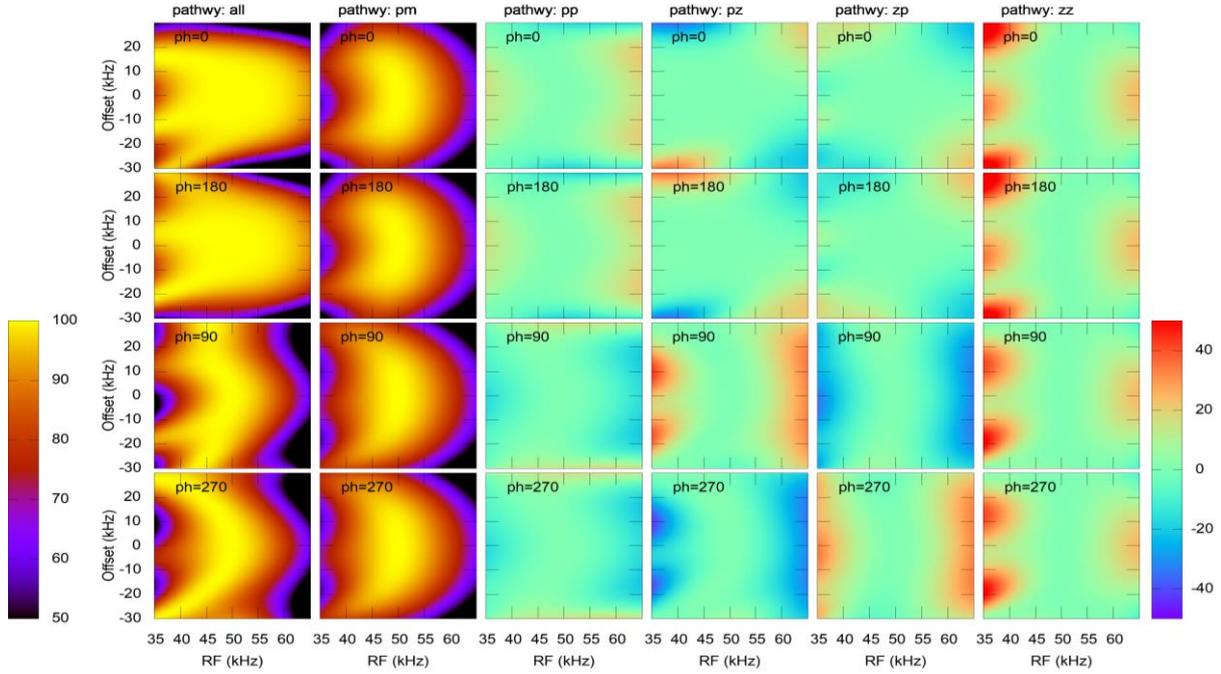

**Fig.S6.** Simulated $E_{rf}$ efficiency versus rf-field and offset for UDEFT using $CP_{x1}$ as refocusing pulse and $AP_z$ as inversion pulse without a selection of coherence transfer pathway (1st column), with coherence orders during the 1st and 2nd $\tau$ delays equal to +1 and −1 (2nd column), +1 and +1 (3rd column), +1 and 0 (4th column), 0 and +1 (5th column), and 0 and 0 (6th column). The phases of the refocusing pulse is 0, 180°, 90°, 270° for rows 1, 2, 3, 4 respectively. The second π/2-pulse phase was reversed when the refocusing pulse phase was 90° or 270°. The UDEFT sequences are: $90_{90}$-$\tau_0$-$(CP_{x1})_0$-$\tau_0$-$90_{90}$-$AP_z$, $90_{90}$-$\tau_0$-$(CP_{x1})_{180}$-$\tau_0$-$90_{90}$-$AP_z$, $90_{90}$-$\tau_{180}$-$(CP_{x1})_{90}$-$\tau_{180}$-$90_{270}$-$AP_z$, and $90_{90}$-$\tau_{180}$-$(CP_{x1})_{270}$-$\tau_{180}$-$90_{270}$-$AP_z$, for the 1st, 2nd, 3rd, and 4th row, respectively, where the subscript after $\tau$ denotes the phase of the receiver. $B_0$ = 9.4 T, $\nu_{1nom}$ = 50 kHz, $\nu_R$ = 10 kHz, CSA = 0.

# VI III. Experimental results

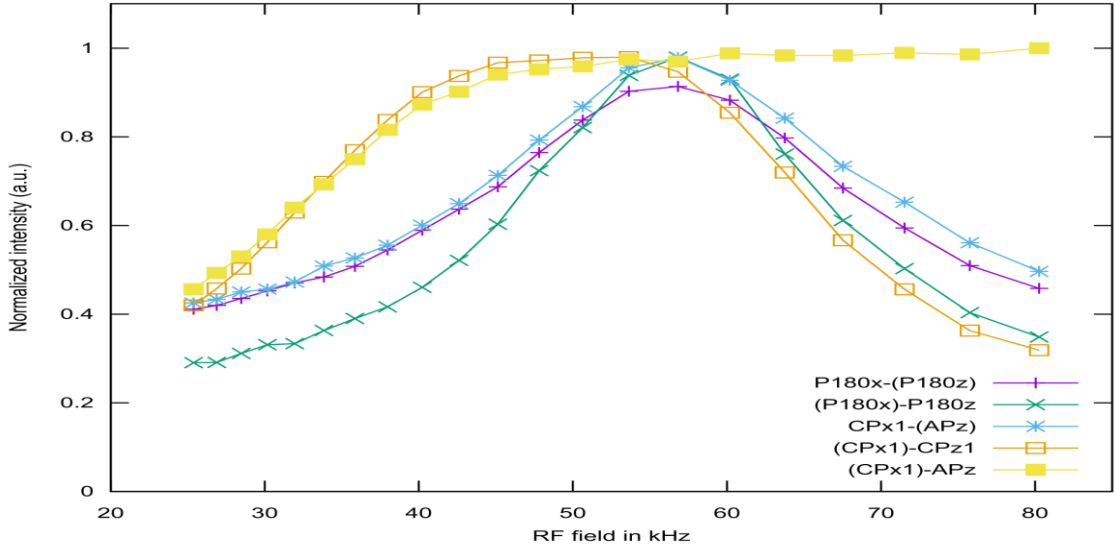

**Fig.S7.** Experimental $^{29}$Si UDEFT signal of $^{29}$Si-enriched amorphous silica sample versus the rf-amplitude of the refocusing and inversion π-pulses. For each such pair one rf-field (that with parentheses) was fixed at $\nu_1$ = 57 kHz (62.5 kHz for $AP_z$), whereas the other was changed about this nominal value. $\nu_1$ = 57 kHz was also used for the π/2-pulses. For $AP_z$, $\tau_p$ = 50 μs and $\Delta\nu_{0,max}$ = 4 MHz. $B_0$ = 9.4 T, $\nu_R$ = 10 kHz, $\tau$ = 2 ms, NS = 16, $\tau_{RD}$ = 1 s.

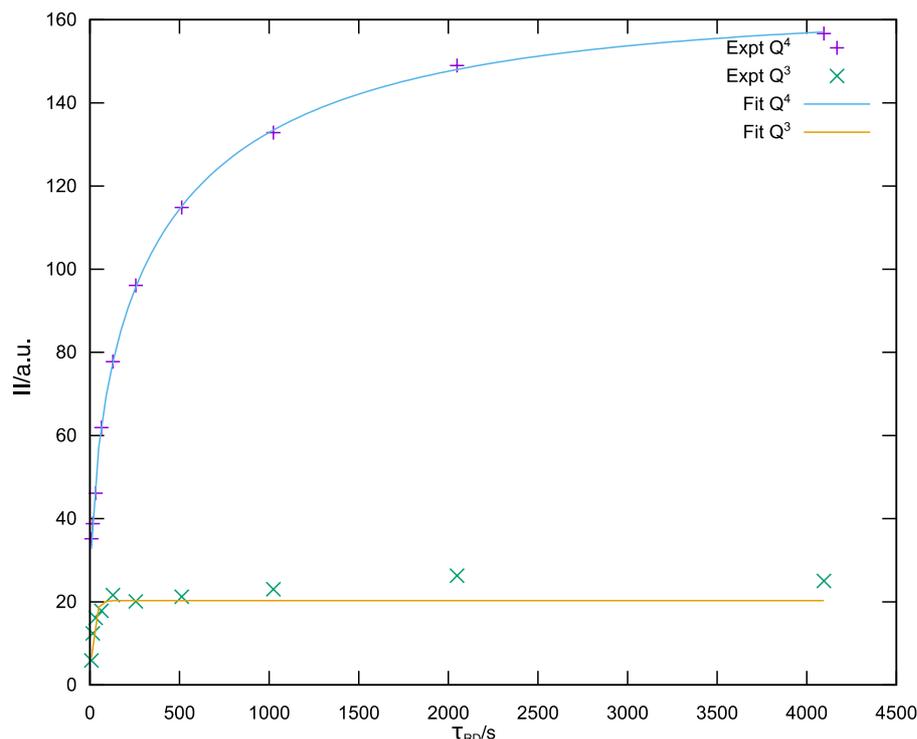

**Fig.S8**. Experimental build-up curves of $^{29}$Si saturation-recovery Q$^3$ and Q$^4$ integrated intensities of SBA-15. The continuous lines are the best-fits of the experimental intensities to a stretched exponential function, $II(\tau) = II_\infty\{1-\exp[(-\tau_{RD}/T_1)^\beta]\}$, where $II_\infty$ is the asymptotic integrated intensity for $\tau \gg T_1$ and $\beta \leq 1$. The best-fit parameters $\{II_\infty, \beta, T_1\}$ for Q$^3$ and Q$^4$ signals are {20, 1, 19 s} and {162, 0.52, 394 s}, respectively. Q$^3$ and Q$^4$ sites were deconvoluted using DMFIT. After saturation and recovery, 128 transients were recorded using UDEFT sequence with a 1s repetition rate.

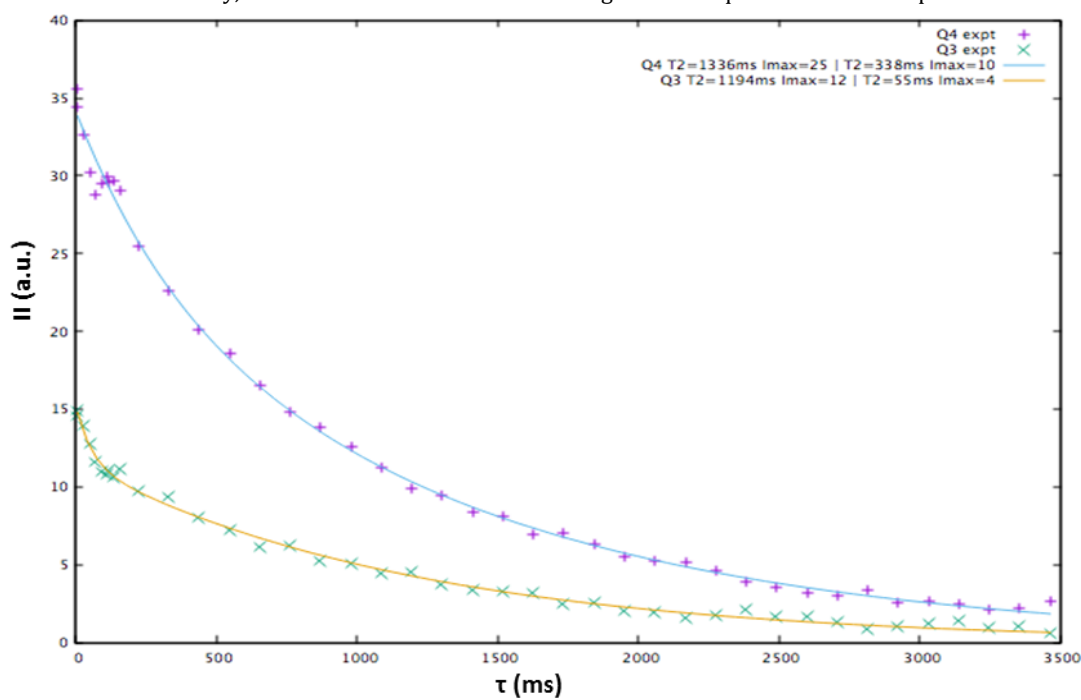

**Fig.S9**. Experimental decay of $^{29}$Si spin-echo Q$^3$ and Q$^4$ integrated intensities (using DMFIT) of SBA-15 versus the echo delay, $\tau$. After the echo sequence, magnetization was detected during a CPMG acquisition. The continuous lines are the best-fits of the experimental intensities to a bi-exponential function $II(\tau) = II_{s,\infty}.\exp(-\tau/T'_{2s}) + II_{f,\infty}.\exp(-\tau/T'_{2f})$ where $II_{i,\infty}$ with i = s or f is the asymptotic integrated intensity for $\tau \gg T_{2i}'$ corresponding to the slow and fast components, respectively. The best-fit parameters $\{II_{f,\infty}, T_{2f}', II_{s,\infty}, T_{2s}'\}$ for Q$^3$ and Q$^4$ signals are {4, 0.05 s, 12, 1.19 s} and {10, 0.34 s, 25, 1.34 s}, respectively.

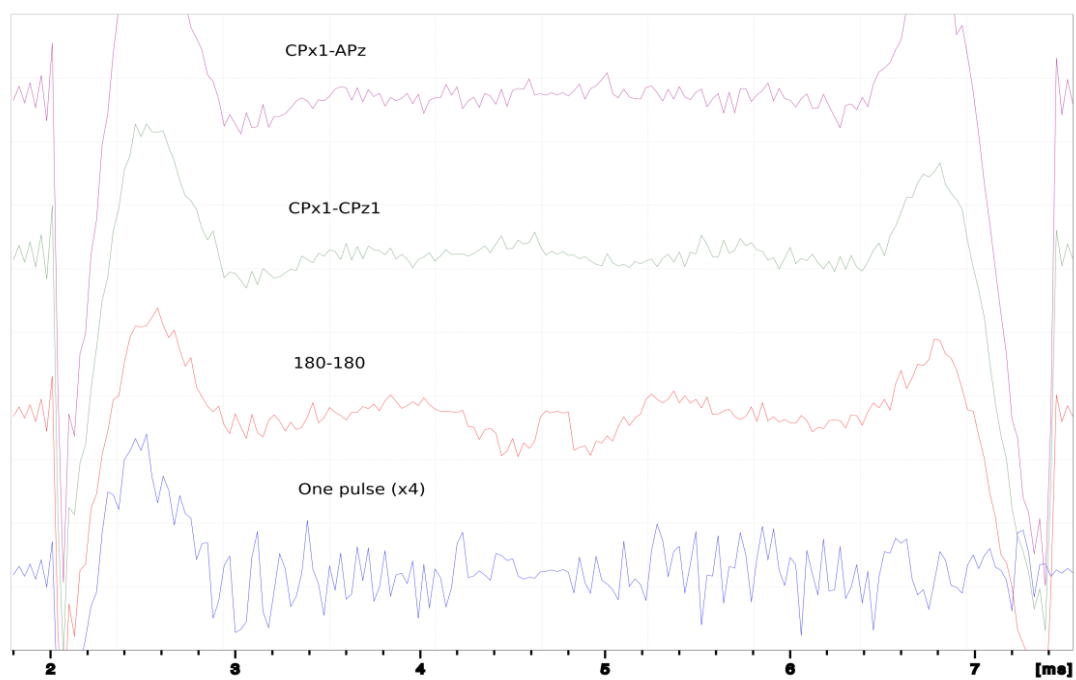

**Fig.S10**. Experimental FIDs of SBA-15 during the two τ delays of UDEFT using $CP_{x1}$-$AP_z$, $CP_{x1}$-$CP_{z1}$ and $P180_x$-$P180_z$. The FID of SP experiment is also displayed during 2τ at the bottom. Experimental parameters are those of Fig.**7**.

| EXP | $D_1$ (s) | D (%) | $T^2$ (%) | $T^3$ (%) | S/N (D) | S/N ($T^3$) |
|---|---|---|---|---|---|---|
| ZG | 180 | 7,9 | 15,3 | 76,8 | 33 | 96 |
| UDEFT | 12,5 | 9,5 | 15,9 | 74,6 | 28 | 62 |
| '' | 25 | 8,5 | 15,8 | 75,7 | 32 | 82 |
| '' | 50 | 8,0 | 15,0 | 77,1 | 38 | 106 |
| '' | 75 | 7,4 | 15,2 | 77,4 | 36 | 106 |
| '' | 100 | 7,1 | 15,3 | 77,7 | 42 | 130 |
| '' | 150 | 6,9 | 15,6 | 77,4 | 38 | 119 |

**Table.S4.** Quantitative analysis of the proportions and S/Ns of the $^{29}Si$ MAS spectra of flame retardant material shown in Fig.**9**.

## IV. Pulse sequences

The pulse sequences for various Bruker consoles are provided in a separate archive.

## V. Simpson input files used for Figs. 4 and S4

The following Simpson input files were used for the Figs.**4** and **S4**. A geometric average must be performed separately to account for the 2 phase cycling.

```
spinsys {
  channels  29Si
  nuclei    29Si
  shift     1 10000 10000 0 0 0 0
}
par {
  proton_frequency  400e6
```

```
    spin_rate        10000
    np               1
    sw               10000
    crystal_file     zcw88
    gamma_angles     7
    start_operator   I1z
    detect_operator  I1z
    variable rf0     50000
    variable rflist  50000
    variable isolist 10000
    variable OneDone 1
    # Adiabatic pulse RF factor:
    # Adiabatic RF is ArfF * RF of nutation pulses
    variable ArfF    1.0
    verbose          0000
#   num_cores        1
}

proc APz {rf phase {preparePhi 0} } {
   global par

   set Tinc 100e-3
   set Tp   50
   set n    [expr $Tp/$Tinc]

if $preparePhi {
set philist ""
set rfilist ""

# other fixed shape pulse parameters
   set rf0AP [expr $par(rf0)*$par(ArfF)]
   set psi  10.0
   set K    atan(30)
   set Q    5

   set pi   [expr atan(1)*4]
# wmax : max frequency offset from rf0AP and Q factor
   set wmax [expr tan($K)*$pi*$rf0AP**2*$Tp*(1e-6)/($K*$Q)]

   if $par(OneDone) {
   # wmax is printed once only
#     puts "Offsetmax=$wmax"
     set par(OneDone) 0
   }

  for {set i 0} {$i< $n} {incr i 1} {
      set x [expr ($i+0.5)/$n]

      if {$i<$n/2} {
        set rfi [expr tanh(2*$psi*$x)]
      } else {
        set rfi [expr tanh(2*$psi*(1-$x))]
      }
      set phi [expr 360*$wmax*$Tp*(1e-6)*log(abs(cos($K*(1-2*$x))))/(2*tan($K)*$K)+$phase]

      set philist "$philist $phi"
      set rfilist "$rfilist $rfi"
   }
   return [list $rfilist $philist]
} #endif

  set rfilist [lindex $par(ilist) 0]
  set philist [lindex $par(ilist) 1]

  for {set i 0} {$i< $n} {incr i 1} {
      pulse $Tinc [expr $rf*$par(ArfF)*[lindex $rfilist $i]] [lindex $philist $i]
   }

   # test OK when printing derivative of phi with time one gets a tanhtan shape with max
offset +-wmax
}

proc Px180 {rf phase {len 0}} {
  global par
  set p180 [expr 1e6/360*180/$par(rf0)]
  if { $len == "length" } {return [expr $p180]}
   pulse $p180 $rf [expr 0+$phase]
```

```
  }
proc CPx1 {rf phase {len 0}} {
   global par
    set p59 [expr 1e6/360*59/$par(rf0)]
    set p298 [expr 1e6/360*298/$par(rf0)]

   if { $len == "length" } {return [expr 2*$p59+$p298]}

    pulse $p59 $rf [expr 180+$phase]
    pulse $p298 $rf [expr 0+$phase]
    pulse $p59 $rf [expr 180+$phase]
}

proc CPx2 {rf phase {len 0}} {
   global par
    set p58 [expr 1e6/360*58/$par(rf0)]
    set p140 [expr 1e6/360*140/$par(rf0)]
    set p344 [expr 1e6/360*344/$par(rf0)]

   if { $len == "length" } {return [expr 2*($p58+$p140)+$p344]}

    pulse $p58 $rf [expr 0+$phase]
    pulse $p140 $rf [expr 180+$phase]
    pulse $p344 $rf [expr 0+$phase]
    pulse $p140 $rf [expr 180+$phase]
    pulse $p58 $rf [expr 0+$phase]
}

proc CPz1 {rf phase {len 0}} {
   global par
    set p90 [expr 1e6/360*90/$par(rf0)]
    set p240 [expr 1e6/360*240/$par(rf0)]
   if { $len == "length" } {return [expr 2*$p90+$p240]}
    pulse $p90 $rf [expr 0+$phase]
    pulse $p240 $rf [expr 90+$phase]
    pulse $p90 $rf [expr 0+$phase]
}

proc CPz2 {rf phase {len 0}} {
   global par
    set p90 [expr 1e6/360*90/$par(rf0)]
    set p180 [expr 1e6/360*180/$par(rf0)]
   if { $len == "length" } {return [expr 2*$p90+$p180]}
    pulse $p90 $rf [expr 90+$phase]
    pulse $p180 $rf [expr 0+$phase]
    pulse $p90 $rf [expr 90+$phase]
}

proc pulseq {} {
   global par
   matrix set 1 operator I1z
   matrix set 2 coherence {{ +1 } {-1 }}

   if {$par(inversion)=="APz"} {
     # prepare philist and amplist stored in ilist
     set par(ilist) [eval APz $par(rf0) 0 1]
   }
    set t90 [expr 0.25e6/$par(rf0)]
    set lenCPx [eval $par(refocus) $par(rf0)  0 length]
    set delEcho  [expr 2e3-($t90+$lenCPx)/2]

   foreach rf $par(rflist) {

    reset
    pulse $t90 $rf 90
    delay $delEcho
    eval $par(refocus) $rf $par(refPhase)
    delay $delEcho
    pulse $t90 $rf 90
    eval $par(inversion) [expr $rf] $par(invPhase)
    acq
    }
}

proc main {} {
   global par
```

```
  set par(inversion) Px180
  set par(refocus) Px180
  set par(rflist) ""
  for {set rf [expr 70000/2.0]} {$rf<=[expr 130000/2.0]} {set rf [expr $rf+2000/2.0]} {
    set par(rflist) "$par(rflist) $rf"
  }
  set par(np) [expr [llength $par(rflist)]]

  set par(isolist) ""
  for {set iso -30000} {$iso<=30000} {set iso [expr $iso+2000]} {
    set par(isolist) "$par(isolist) $iso"
  }
  set par(csalist) ""
  for {set csa 0} {$csa<=40000} {set csa [expr $csa+2000]} {
    set par(csalist) "$par(csalist) $csa"
  }

          foreach csa $par(csalist) {
  foreach par(refocus) { Px180 CPx1 CPx2} {
    foreach par(inversion) {Px180 CPz1 CPz2 APz } {
      foreach par(refPhase) { 0 180 } {
         foreach par(invPhase) { 0 180 90 270 } {
            set res_name "$par(name)-$par(refocus)_$par(refPhase)-$par(inversion)_$par(invPhase)-csa=$csa.res"
            puts $res_name
            set File [open $res_name w]
            foreach iso $par(isolist) {
              set f [fsimpson [list [list shift_1_iso $iso] [list shift_1_aniso $csa] ]]
              for {set i 1} {$i <= $par(np)} {incr i} {
                 set Sr [expr [findex $f [expr $i] -re ] /0.5 *100]
                 set Si [expr [findex $f [expr $i] -im ] /0.5 *100]
                 puts $File "[expr [lindex $par(rflist) [expr $i-1]]/1000] [expr $iso/1000] $Sr $Si"
              }
              puts $File ""
              funload $f
            }
            close $File
          }
        }
      }
    }
  }
}
```